\documentclass[letterpaper,11pt]{article}
\usepackage[dvips]{graphics,color}
\usepackage{latexsym}
\usepackage{amsmath}
\usepackage{amssymb}
\usepackage{amsfonts,amsthm}
\usepackage[toc,page]{appendix}

\newtheorem{theorem}{Theorem}[section]

\newcommand{\nca}{\mbox{\it nca}}

\newcommand{\ignore}[1]{}

\makeatletter
\def\argmin{\mathop{\operator@font argmin}}
\makeatother

\newcommand{\reach}{\rightsquigarrow}

\begin{document}

\title{Join-Reachability Problems in Directed Graphs\thanks{This research project has been funded by the John S. Latsis Public Benefit Foundation. The sole responsibility for the content of this paper lies with its authors.}}

\author{Loukas Georgiadis$^1$ \and Stavros D. Nikolopoulos$^2$ \and Leonidas Palios$^2$}


\maketitle

\begin{abstract}
For a given collection $\mathcal{G}$ of directed graphs we define the \emph{join-reachability graph} of $\mathcal{G}$, denoted by $\mathcal{J}(\mathcal{G})$, as the directed graph that, for any pair of vertices $a$ and $b$, contains a path from $a$ to $b$ if and only if such a path exists in all graphs of $\mathcal{G}$. Our goal is to compute an efficient representation of  $\mathcal{J}(\mathcal{G})$. In particular, we consider two versions of this problem. In the \emph{explicit} version we wish to construct the smallest join-reachability graph for $\mathcal{G}$. In the \emph{implicit} version we wish to build an efficient data structure (in terms of space and query time) such that we can report fast the set of vertices that reach a query vertex in all graphs of $\mathcal{G}$. This problem is related to the well-studied \emph{reachability problem} and is motivated by emerging applications of graph-structured databases and graph algorithms. We consider the construction of join-reachability structures for two graphs and develop techniques that can be applied to both the explicit and the implicit problem. First we present optimal and near-optimal structures for paths and trees. Then, based on these results, we provide efficient structures for planar graphs and general directed graphs.
\end{abstract}

\footnotetext[1]{Department of Informatics and Telecommunications Engineering, University of Western Macedonia, Kozani, Greece. E-mail: {\tt lgeorg@uowm.gr}.}
\footnotetext[2]{Department of Computer Science, University of Ioannina, Ioannina, Greece. E-mail: {\tt \{stavros,palios\}@cs.uoi.gr}.}

\section{Introduction}

In the \emph{reachability problem} our goal is to preprocess a (directed or undirected) graph $G$ into a data structure that can quickly answer queries that ask if a vertex $b$ is reachable from a vertex $a$. This problem has numerous and diverse applications, including internet routing, geographical navigation, and knowledge-representation systems~\cite{reachability:WHYYY06}. Recently, the interest in graph reachability problems has been rekindled by emerging applications of graph data structures in areas such as the semantic web, bio-informatics and social networks. These developments together with recent applications in graph algorithms~\cite{fdom:G,2-vc:g,domv:gt05} have motivated us to introduce the study of the \emph{join-reachability problem} that we define as follows: We are given a collection $\mathcal{G}$ of $\lambda$ directed graphs $G_i=(V_i,A_i)$, $1 \le i \le \lambda$, where each graph $G_i$ represents a binary relation $R_i$ over a set of elements $V \subseteq V_i$ in the following sense: For any $a,b \in V$, we have $a R_i b$ if and only if $b$ is reachable from $a$ in $G_i$. Let $\mathcal{R} \equiv \mathcal{R}(\mathcal{G})$ be the binary relation over $V$ defined by: $a \mathcal{R} b$ if and only if $a R_i b$ for all $i \in \{1,\ldots,\lambda\}$ (i.e., $b$ is reachable from $a$ in all graphs in $\mathcal{G}$). We can view $\mathcal{R}$ as a type of {\sc join} operation on graph-structured databases. Our objective is to find an efficient representation of this relation. To the best of our knowledge, this problem has not been previously studied. We will restrict our attention to the case of two input graphs ($\lambda=2$).

\paragraph{Contribution.} In this paper we explore two versions of the join-reachability problem. In the \emph{explicit} version
we wish to represent $\mathcal{R}$ with a directed graph $\mathcal{J} \equiv \mathcal{J}(\mathcal{G})$, which we call the \emph{join-reachability graph of} $\mathcal{G}$, i.e., for any $a,b \in V$, we have $a \mathcal{R} b$ if and only if $b$ is reachable from $a$ in $\mathcal{J}$. Our goal is to minimize the size (i.e., the number of vertices plus arcs) of $\mathcal{J}$. We consider this problem in Sections \ref{sec:complexity} and \ref{sec:combinatorial}, and present results on the computational and combinatorial complexity of $\mathcal{J}$.
In the \emph{implicit} version we wish to represent $\mathcal{R}$
with an efficient data structure (in terms of space and query time) that can report fast all elements $a \in V$ satisfying $a \mathcal{R} b$ for any query element $b \in V$.
We deal with the implicit problem in Section \ref{sec:ds}. First we
describe efficient join-reachability structures for simple graph classes. Then, based on these results, we consider planar graphs and general directed graphs. Also, in Appendix \ref{appendix:planar-st} and Appendix \ref{appendix:lattice} we consider join-reachability structures for planar st-graphs and lattices. Although we focus on the case of two directed graphs ($\lambda=2$), we note that some of our results are easily extended for $\lambda \ge 3$ with the use of appropriate multidimensional geometric structures.

\paragraph{Applications.}
Instances of the join-reachability problem appear in various applications. For example, in the rank aggregation problem~\cite{rank-aggregation:ckns01} we are given a collection of rankings of some elements and we may wish to report which (or how many) elements  have the same ranking relative to a given element. This is a special version of join-reachability since the given collection of rankings can be represented by a collection of directed paths with the elements being the vertices of the paths. Similarly, in a graph-structured database with an associated ranking of its vertices we may wish to find the vertices that are related to a query vertex and have higher or lower ranking than this vertex. Instances of join-reachability also appear in graph algorithms arising from program optimization. Specifically, in \cite{fdom:G} we need a data structure capable of reporting which vertices satisfy certain ancestor-descendant relations in a collection of rooted trees. \ignore{Also, in current work in progress, we show that join-reachability structures for two trees can yield efficient solutions to special cases of the interprocedural dominance problem~\cite{inter-dom:ds}.}
Moreover, in \cite{domv:gt05} it is shown that any directed graph $G$ with a distinguished source vertex $s$ has two spanning trees rooted at $s$ such that a vertex $a$ is a dominator of a vertex $b$ (meaning that all paths in $G$ from $s$ to $b$ pass through $a$) if and only if $a$ is an ancestor of $b$ in both spanning trees. This generalizes the graph-theoretical concept of \emph{independent spanning trees}. Two spanning trees of a graph $G$ are independent if they are both rooted at the same vertex $r$ and for each vertex $v$ the paths from $r$ to $v$ in the two trees are internally vertex disjoint. Similarly, $\lambda$ spanning trees of $G$ are independent if they are pairwise independent. In this setting, we can apply a join-reachability structure to decide if $\lambda$ given spanning trees are independent. Finally we note that a variant of the join-reachability problem we defined here appears in the context of a recent algorithm for computing two internally vertex-disjoint paths for any pair of query vertices in a 2-vertex connected directed graph~\cite{2-vc:g}.

\paragraph{Preliminaries and Related Work.}
The reachability problem is easy in the undirected case since it suffices to compute the connected components of the input graph. Similarly, the undirected version of the join-reachability problem is also easy, as given the connected components of two undirected graphs $G_1$ and $G_2$ with $n$ vertices, we can compute the connected components of $\mathcal{J}(\{G_1,G_2\})$ in $O(n)$ time. On the other hand, no reachability data structure is currently known to simultaneously achieve $o(n^2)$ space and $o(n)$ query time for a general directed graph with $n$ vertices~\cite{reachability:WHYYY06}.
Nevertheless, efficient reachability structures do exist for several important cases. First, asymptotically optimal structures exist for rooted trees~\cite{transitive-db:abj} and planar directed graphs with one source and one sink~\cite{planar-reach:kameda,planar-reach:tt}. For general planar graphs Thorup~\cite{planar-reach:thorup} gives an $O(n \log n)$-space structure with constant query time.
Talamo and Vocca~\cite{lattice-reach:tv99} achieve constant query time
for lattice partial orders with an $O(n \sqrt{n})$-space structure.

\paragraph{Notation.}
In the description of our results we use the following notation and terminology. We denote the vertex set and the arc set of a directed graph (digraph) $G$ by $V(G)$ and $A(G)$, respectively. Without loss of generality we assume that $V(G)=V$ for all $G \in \mathcal{G}$. The size of $G$, denoted by $|G|$, is equal to the number of arcs plus vertices, i.e., $|G|=|V|+|E|$. We use the notation $a \rightsquigarrow_{G} b$ to denote that $b$ is reachable from $a$ in $G$. (By definition $a \rightsquigarrow_{G} a$ for any $a \in V$.) The \emph{predecessors} of a vertex $b$ are the vertices that reach $b$, and the \emph{successors} of a vertex $b$ are the vertices that are reached from $b$.
Let $P$ be a directed path (dipath); the \emph{rank} of $a \in P$, $r_P(a)$, is equal to the number of predecessors of $a$ in $P$ minus one, and the \emph{height} of $a \in P$, $h_P(a)$, is equal to the number of successors of $a$ in $P$ minus one.
For a rooted tree $T$, we let $T(a)$ denote the subtree rooted at $a$ and let $\nca_{T}(a,b)$ denote the nearest common ancestor of $a$ and $b$.
We will deal with two special types of directed rooted trees: In an \emph{in-tree}, each vertex has exactly one outgoing arc except for the root which has none; in an \emph{out-tree}, each vertex has exactly one incoming arc except for the root which has none. We use the term \emph{unoriented tree} for a directed tree with no restriction on the orientation of its arcs. Similarly, we use the term \emph{unoriented dipath} to refer to a path in the undirected sense, where the arcs can have any orientation.
In our constructions we map the vertices of $V$ to objects in a $d$-dimensional space and use the notation $x_i(a)$ to refer to the $i$th coordinate that vertex $a$ receives.
Finally, for any two vectors $\xi=(\xi_1,\ldots,\xi_d)$ and $\zeta=(\zeta_1,\ldots,\zeta_d)$, the notation $\xi \le \zeta$ means that $\xi_i \le \zeta_i$ for $i=1,\ldots,d$.

\subsection{Preprocessing: Computing Layers and Removing Cycles}
\label{sec:preprocessing}

\paragraph{Thorup's Layer Decomposition.}
In \cite{planar-reach:thorup} Thorup shows how to reduce the reachability problem for any digraph $G$ to reachability in some digraphs with special properties, called \emph{2-layered digraphs}. A \emph{t-layered spanning tree} $T$ of $G$ is a rooted directed tree such that any path in $T$ from the root (ignoring arc directions) is the concatenation of at most $t$ dipaths in $G$. A digraph $G$ is \emph{t-layered} if it has such a spanning tree.
Now we provide an overview of Thorup's reduction. The vertices of $G$ are partitioned into layers $L_0, L_1, \ldots, L_{\mu-1}$ that define a sequence of digraphs $G^0, G^1, \ldots, G^{\mu-1}$ as follows. An arbitrary vertex $v_0 \in V(G)$ is chosen as a root. Then, layer $L_0$ contains $v_0$ and the vertices that are reachable from $v_0$. For odd $i$, layer $L_i$ contains the vertices that reach the previous layers $L_j$, $j<i$. For even $i$, layer $L_i$ contains the vertices that are reachable from the previous layers $L_j$, $j<i$. To form $G^i$ for $i>0$ we contract the vertices in layers $L_j$ for $j \le i-1$ to a single root vertex $r_0$; for $i=0$ we set $r_0=v_0$. Then $G^i$ is induced by $L_{i}$,  $L_{i+1}$ and $r_0$.
It follows that each $G^i$ is a 2-layered digraph.
Let $\iota(v)$ denote the index of the layer containing $v$, that is, $\iota(v) = i$ if and only if $v \in L_i$. The key properties of the decomposition are: (i) all the predecessors of $v$ in $G$ are contained in $G^{\iota(v)-1}$ and $G^{\iota(v)}$, and (ii) $\sum_{i} |G^{i}| = O(|G|)$.

\paragraph{Removing Cycles.}
In the standard reachability problem, a useful preprocessing step that can reduce the size of the input digraph is to contract its strongly connected components (strong components) and consider the resulting acyclic graph. When we apply the same idea to join-reachability we have to deal with the complication that the strong components in the two digraphs may differ. Still, we can construct two acyclic digraphs $\hat{G}_1$ and $\hat{G}_2$ such that, for any $a,b \in V$, $a \rightsquigarrow_{\mathcal{J}(\{G_1,G_2\})} b$ if and only if $a \rightsquigarrow_{\mathcal{J}(\{\hat{G}_1,\hat{G}_2\})} b$, and $|\hat{G}_i| \le |G_i|$, $i=1,2$. This is accomplished as follows. First, we compute the strong components of $G_1$ and $G_2$ and order them topologically. Let $G'_i$, $i=1,2$, denote the digraph produced after contracting the strong components of $G_i$. (We remove loops and duplicate arcs so that each $G'_i$ is a simple digraph.) Also, let $C_i^j$ denote the $j$th strong component of $G_i$. We partition each component $C_i^j$ into subcomponents such that two vertices are in the same subcomponent if and only if they are in the same strong component in both $G_1$ and $G_2$. The subcomponents are the vertices of $\hat{G}_1$ and $\hat{G}_2$. Next we describe how to add the appropriate arcs. The process is similar for the two digraphs so we consider only $\hat{G}_1$.

\begin{figure}[t]
\begin{center}
\scalebox{0.55}[0.55]{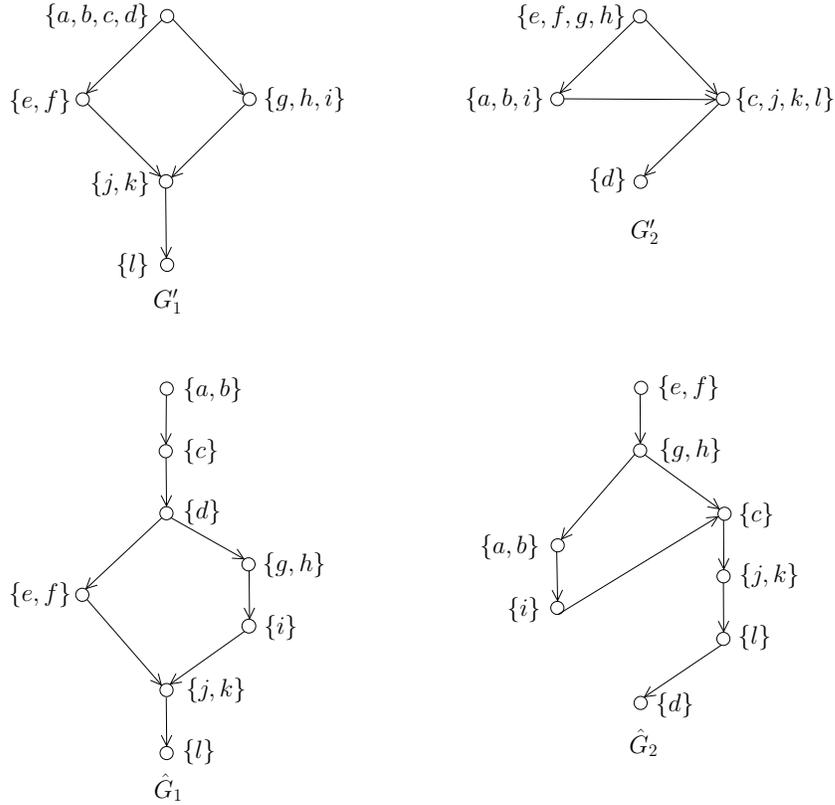}
\end{center} \caption{The contracted digraphs $G'_1$ and $G'_2$ and their corresponding acyclic digraphs $\hat{G}_1$ and $\hat{G}_2$.\label{fig:strong-components}}
\end{figure}

Let $C_1^{j,1}, C_1^{j,2}, \ldots, C_1^{j,l_j}$ be the subcomponents of $C_1^{j}$, which are ordered with respect to the topological order of $G'_2$. That is, if $x \in C_1^{j,i}$ and $y \in C_1^{j,i'}$, where $i<i'$, then in the topological order of $G'_2$ the component of $x$ precedes the component of $y$. We connect the subcomponents by adding the arcs $(C_1^{j,i},C_1^{j,i+1})$ for $1 \le i < l_j$. Moreover, for each arc $(C_1^{i},C_1^{j})$ in $A(G'_1)$ we add the arc $(C_1^{i,l_i},C_1^{j,1})$ to $A(\hat{G}_1)$, where $C_1^{i,l_i}$ is the last subcomponent of $C_1^{i}$. See Figure \ref{fig:strong-components}. It is straightforward to verify that $a \reach_\mathcal{J} b$ if and only if $a$ and $b$ are in the same subcomponent or the subcomponent of $a$ is a predecessor of the subcomponent of $b$ in both $\hat{G}_1$ and $\hat{G}_2$.

\section{Computational Complexity of Computing the Smallest $\mathcal{J}(\{G_1,G_2\})$}
\label{sec:complexity}

We explore the computational complexity of computing the smallest $\mathcal{J}(\{G_1,G_2\})$: Given two digraphs $G_1=(V,A_1)$ and $G_2=(V,A_2)$ we wish to compute a digraph $\mathcal{J} \equiv \mathcal{J}(\{G_1,G_2\})$ of minimum size such that for any $a,b \in V$, $a \reach_\mathcal{J} b$ if and only if $a \reach_{G_1} b$ and $a \reach_{G_2} b$. We consider two versions of this problem, depending on whether $\mathcal{J}$ is allowed to have Steiner vertices (i.e., vertices not in $V$) or not: In the \emph{unrestricted} version $V(\mathcal{J}) \supseteq V$, while in the \emph{restricted} version $V(\mathcal{J}) = V$.
Computing $\mathcal{J}$ is NP-hard in the unrestricted case. This is implied by a straightforward reduction to the \emph{reachability substitute problem}, which was shown to be NP-hard by Katriel et al.~\cite{reach-sub:KKS}. In this problem we are given a digraph $H$ and a subset $U \subseteq V(H)$, and ask for the smallest digraph $H^{\ast}$ such that for any $a, b \in U$, $a \reach_{H^{\ast}} b$ and only if $a \reach_{H} b$.
For the reduction, we let $G_1=H$ and let $G_2$ contain all the arcs connecting vertices in $U$ only, that is, $A(G_2)=U \times U$. Clearly, for any $a,b \in U$ we have $a \reach_\mathcal{J} b$ if and only if $a \reach_H b$. Therefore computing the smallest join-reachability graph is equivalent to computing $H^{\ast}$. In the restricted case, on the other hand, we can compute $\mathcal{J}$ using transitive closure and transitive reduction computations, which can be done in polynomial time~\cite{transitive-reduction:agu}. (This is done as follows: First we compute the transitive closure matrices $M_1$ and $M_2$ of $G_1$ and $G_2$ respectively. Then we form the transitive closure matrix $M$ of $\mathcal{J}$ by taking the \textsc{and}
operation of corresponding entries in $M_1$ and $M_2$. Finally we compute the transitive reduction of the resulting transitive closure matrix $M$.) This implies the next theorem.
\begin{theorem}
\label{thm:computational}
Let $\mathcal{J}$ be the smallest join-reachability graph of a collection of digraphs. The computation of $\mathcal{J}$ is feasible in polynomial time if Steiner vertices are not allowed, and NP-hard otherwise.
\end{theorem}
The existence of Steiner vertices can reduce the size of $\mathcal{J}$ significantly.
Consider for example a complete bipartite digraph $G$ with  $V(G)=X \cup Y$ and $A(G) = X \times Y$.
This digraph has the same transitive closure as the digraph $G'$ with $V(G')=V(G)\cup\{z\}$ and $A(G')=\{(x,z), (z,y) \ | \ x \in X, y \in Y\}$. In Section \ref{sec:combinatorial} we explore the combinatorial complexity of the unrestricted join-reachability graph and provide bounds for $|\mathcal{J}|$ in several cases.

\section{Combinatorial Complexity of $\mathcal{J}(\{G_1,G_2\})$}
\label{sec:combinatorial}

In this section we provide bounds on the size of $\mathcal{J}(\{G_1,G_2\})$ in several cases. These results are summarized in the next theorem.
\begin{theorem}
\label{thm:combinatorial}
Given two digraphs $G_1$ and $G_2$ with $n$ vertices, the following bounds on the size of the join-reachability graph $\mathcal{J}(\{G_1,G_2\})$ hold:
\begin{itemize}
\ignore{\item[(a)] $\Theta(n \log n)$ in the worst case when $G_1$ and $G_2$ are dipaths.\ignore{\vspace{-2ex}}}
\item[(a)] $\Theta(n \log n)$ in the worst case when $G_1$ is an unoriented tree and $G_2$ is an unoriented dipath.\ignore{\vspace{-2ex}}
\item[(b)] $O(n \log^2 n)$ when both $G_1$ and $G_2$ are unoriented trees.\ignore{\vspace{-2ex}}
\item[(c)] $O(n \log^2 n)$ when $G_1$ is a planar digraph and $G_2$ is an unoriented dipath.\ignore{\vspace{-2ex}}
\item[(d)] $O(n \log^3 n)$ when both $G_1$ and $G_2$ are planar digraphs.\ignore{\vspace{-2ex}}
\item[(e)] $O(\kappa_1 n \log n)$ when $G_1$ is a digraph that can be covered with $\kappa_1$ vertex-disjoint dipaths and $G_2$ is an unoriented dipath.\ignore{\vspace{-2ex}}
\item[(f)] $O(\kappa_1 n  \log^2 n)$ when $G_1$ is a digraph that can be covered with $\kappa_1$ vertex-disjoint dipaths and $G_2$ is a planar graph.\ignore{\vspace{-2ex}}
\item[(g)] $O(\kappa_1 \kappa_2 n  \log n)$ when each $G_i$, $i=1,2$, is a digraph that can be covered with $\kappa_i$ vertex-disjoint dipaths.
\end{itemize}
\end{theorem}
In the following sections we prove Theorem \ref{thm:combinatorial}. In each case we provide a construction of the corresponding join-reachability graph that achieves the claimed bound. In Section \ref{sec:ds} we provide improved space bounds for the implicit representation of $\mathcal{J}(\{G_1,G_2\})$, i.e., data structures that answer join-reachability reporting queries fast. Still, a process that computes an explicit representation of $\mathcal{J}(\{G_1,G_2\})$ can be useful, as it provides a natural way to handle collections of more than two digraphs (i.e., it allows us to combine the digraphs one pair at a time).

\subsection{Two Paths}
\label{sec:comb:two-paths}

We start with the simplest case where $G_1$ and $G_2$ are dipaths with $n$ vertices. First we show that we can construct a join-reachability graph of size $O(n \log n)$. Given this result we can provide bounds for trees, planar digraphs, and general digraphs. Then we show this bound is tight, i.e., there are instances for which $\Omega(n \log n)$ size is needed.
We begin by mapping the vertices of $V$ to a two-dimensional rank space: Each vertex $a$ receives coordinates $(x_1(a),x_2(a))$ where $x_1(a)=r_{G_1}(a)$ and $x_2(a)=r_{G_2}(a)$. Note that these ranks are integers in the range $[0,n-1]$.  Now we can view these vertices as lying on an $n \times n$ grid, such that each row and each column of the grid contains exactly one vertex.
Clearly, $a \mathcal{R} b$ if and only if $(x_1(a),x_2(a)) \le (x_1(b),x_2(b))$.

\begin{figure}
\begin{center}
\scalebox{0.75}[0.75]{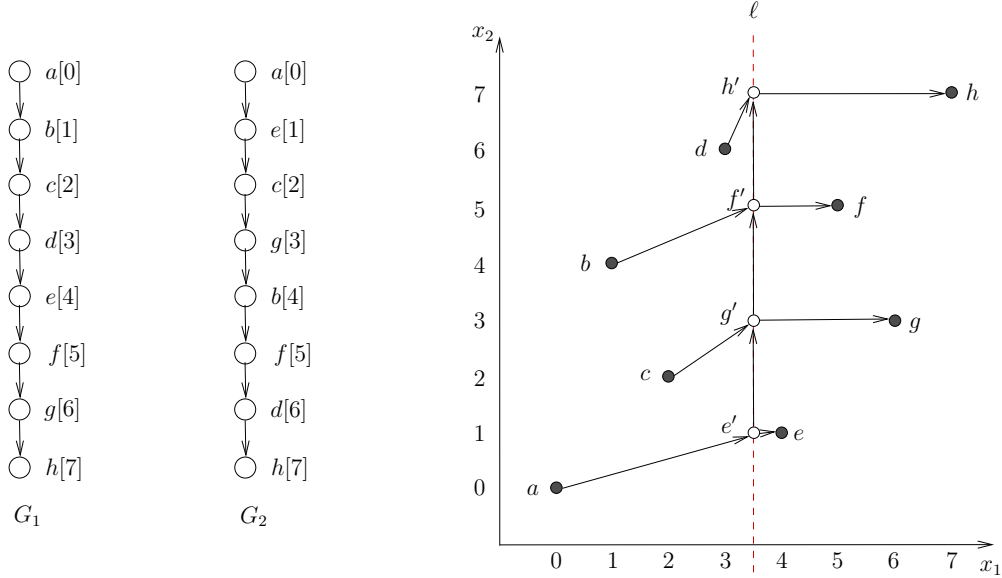}
\end{center}
\caption{The mapping of the vertices of two dipaths to 2d rank space and the construction of $\mathcal{J}_{\ell}$; Steiner vertices in  $\mathcal{J}_{\ell}$ are white. \label{fig:paths}}
\end{figure}

\paragraph{Upper bound.} We use a simple divide-and-conquer method. Let $\ell$ be the vertical line with $x_1$-coordinate equal to $n/2$. A vertex $z$ is \emph{to the right of} $\ell$ if $x_1(z) \ge n/2$ and \emph{to the left of} $\ell$ otherwise. The first step is to construct a subgraph $\mathcal{J}_{\ell}$ of $\mathcal{J}$ that connects the vertices to the left of $\ell$ to the vertices to the right of $\ell$.
For each vertex $b$ to the right of $\ell$ we create a Steiner vertex $b'$ and add the arc $(b',b)$. Also, we assign to $b'$ the coordinates $(n/2,x_2(b))$. We connect these Steiner vertices in a dipath starting from the vertex with the lowest $x_2$-coordinate. Next, for each vertex $a$ to the left of $\ell$ we locate the Steiner vertex $b'$ with the smallest $x_2$-coordinate such that $x_2(a) \le x_2(b')$. If $b'$ exists we add the arc $(a,b')$. See Figure \ref{fig:paths}.
Finally we recurse for the vertices to the left of $\ell$ and for the vertices to the right of $\ell$.
It is easy to see that $\mathcal{J}$ contains a path from $a$ to $b$ if and only if $(x_1(a),x_2(a)) \le (x_1(b),x_2(b))$. To bound $|\mathcal{J}|$ note that we have $O(\log n)$ levels of recursion, and at each level the number of added Steiner vertices and arcs is $O(n)$. Hence, the $O(n \log n)$ bound for two dipaths follows.

The case of two unoriented dipaths $G_1$ and $G_2$ can be reduced to that of dipaths, yielding the same $O(n \log n)$ bound.
\ignore{as stated in Theorem \ref{thm:combinatorial}(b)}This is accomplished by splitting $G_1$ and $G_2$ to maximal subpaths that consist of arcs with the same orientation.
Then $\mathcal{J}$ is formed from the union of separate join-reachability graphs for each pair of subpaths of $G_1$ and $G_2$.
The $O(n \log n)$ bound follows from the fact that each vertex appears in at most two subpaths of each unoriented dipath, so in at most four subgraphs.
\ignore{We omit the details which are similar to the more complicated construction of Section \ref{sec:comb:two-trees}.}
We remark that our construction can be generalized to handle more dipaths, with an $O(\log{n})$ factor blowup per additional dipath.

\ignore{
The above result implies an $O(n \log^2 n)$ bound for the size of the join-reachability graph when $G_1$ is a rooted tree and an $O(n \log^3 n)$
bound when both $G_1$ and $G_2$ are rooted trees, since these cases correspond to collections of three and four dipaths, respectively. (See Section \ref{sec:path-tree}.)
More generally, using the separator dipath decomposition of Thorup~\cite{planar-reach:thorup} we can get the same bounds for the case where $G_1$ is a planar digraph (see Section \ref{sec:planar}).
Similar bounds also hold when $G_1$ is a digraph with a small path cover. More precisely, if $G_1$ can be decomposed into $\kappa$ disjoint paths then we have an additional $O(n \kappa)$ term (see Section \ref{sec:general}). Finally, we note that the data structure of Talamo and Vocca for representing lattices implies an $O(n \sqrt{n})$ bound when $G_1$ is a lattice (see Appendix \ref{appendix:lattice}).
}

\begin{figure}[t]
\begin{center}
\scalebox{0.45}[0.45]{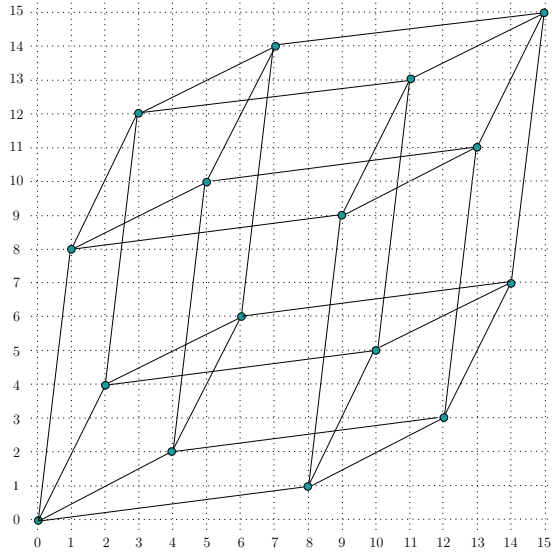}
\end{center}
\caption{The digraph used in the lower bound proof of Section \ref{sec:complexity} for $n=16$. The arcs are directed towards northeast. The $x_2$-coordinate of each vertex is produced by reversing the bits of its $x_1$-coordinate. \label{fig:bit-reversal}}
\end{figure}

\paragraph{Lower bound.} \ignore{The $\Omega(n \log n)$ bound results from a rank space that was previously used in \cite{range-search-lb:chazelle,partial-sums:yao}.}Let $G_1$ be any dipath, and let $x_1(a)=r_{G_1}(a)$. Also let $x_1^i(a)$ denote the $i$th bit in the binary representation of $x_1(a)$ and let $\beta=\lceil \log_2 n \rceil$ be the number of bits in this representation. We use similar notation for $x_2(a)$. We define $G_2$ such that the rank of $a$ in $G_2$ is $x_2(a)=x_1(a)^R$, where $x_1(a)^R$ is the integer formed by the bit-reversal in the binary representation of $x_1(a)$, i.e., $x_2^{i}(a) = x_1^{\beta-1-i}(a)$ for $0 \le i \le \beta-1$.
Let $\mathcal{P}$ be the set that contains all pairs of vertices $(a,b)$ that satisfy $x_1^i(a)=0$, $x_1^i(b)=1$ and $x_1^j(a)=x_1^j(b), j \neq i$,
for $0 \le i \le \beta-1$. Notice that for a pair $(a,b) \in \mathcal{P}$, $x_1(a) < x_1(b)$ and $x_1(a)^R < x_1(b)^R$.
Hence $(x_1(a),x_2(a)) < (x_1(b),x_2(b))$, which implies $a \reach_{\mathcal{J}} b$.
Now let $G$ be the digraph that is formed by the arcs $(a,b) \in \mathcal{P}$. See Figure \ref{fig:bit-reversal}.
Then $a \reach_{G} b$ only if $a \reach_{\mathcal{J}} b$. Moreover, the transitive reduction of $G$ is itself and has size $\Omega(n \log n)$.
We also observe that any two vertices in $G$ share at most one immediate successor.
Therefore the size of $G$ cannot be reduced by introducing Steiner vertices.
This implies that size of $\mathcal{J}$ is also $\Omega(n \log n)$.

\subsection{Tree and Path}
\label{sec:comb:path-tree}

Let $G_1$ be a rooted (in- or out-)tree and $G_2$ a dipath. First we note that the ancestor-descendant relations in a rooted tree can be described by two linear orders (corresponding to a preorder and a postorder traversal of the tree) and therefore we can get an $O(n \log^2{n})$ bound on the size of $\mathcal{J}$ using the result of Section \ref{sec:comb:two-paths}. Here we provide an $O(n \log{n})$ bound, which also holds when $G_1$ is unoriented. This upper bound together with the $\Omega(n \log n)$ lower bound of Section \ref{sec:comb:two-paths} implies Theorem \ref{thm:combinatorial}(a).

Let $T$ be the rooted tree that results from $G_1$ after removing arc directions. We associate each vertex $x \in T$ with a label $h(x)=h_{G_2}(x)$, the height of $x$ in $G_2$.
If $G_1$ is an out-tree then any vertex $b$ must be reachable from all its ancestors $a$ in $T$ with $h(a)>h(b)$. Similarly, if $G_1$ is an in-tree then any vertex $b$ must be reachable from all its descendants $a$ in $T$ with $h(a)>h(b)$. We begin by assigning a depth-first search interval to each vertex in $T$. Let $I(a)=[s(a),t(a)]$ be the interval of a vertex  $a \in T$; $s(a)$ is the time of the first visit to $a$ (during the depth-first search) and $t(a)$ is the time of the last visit to $a$. These times are computed by incrementing a counter after visiting or leaving a vertex during the search. This way all the $s()$ and $t()$ values that are assigned are distinct and for any vertex $a$ we have $1 \le s(a) < t(a) \le 2n$. Moreover, by well-known properties of depth-first search, we have that $a$ is an ancestor of $b$ in $T$ if and only if $I(b) \subseteq I(a)$; if $a$ and $b$ are unrelated in $T$ then $I(a)$ and $I(b)$ do not intersect. Now we map each vertex $a$ to the $x_1$-axis-parallel segment $S(a)=I(a) \times h(a)$.

\begin{figure}
\begin{center}
\scalebox{0.7}[0.7]{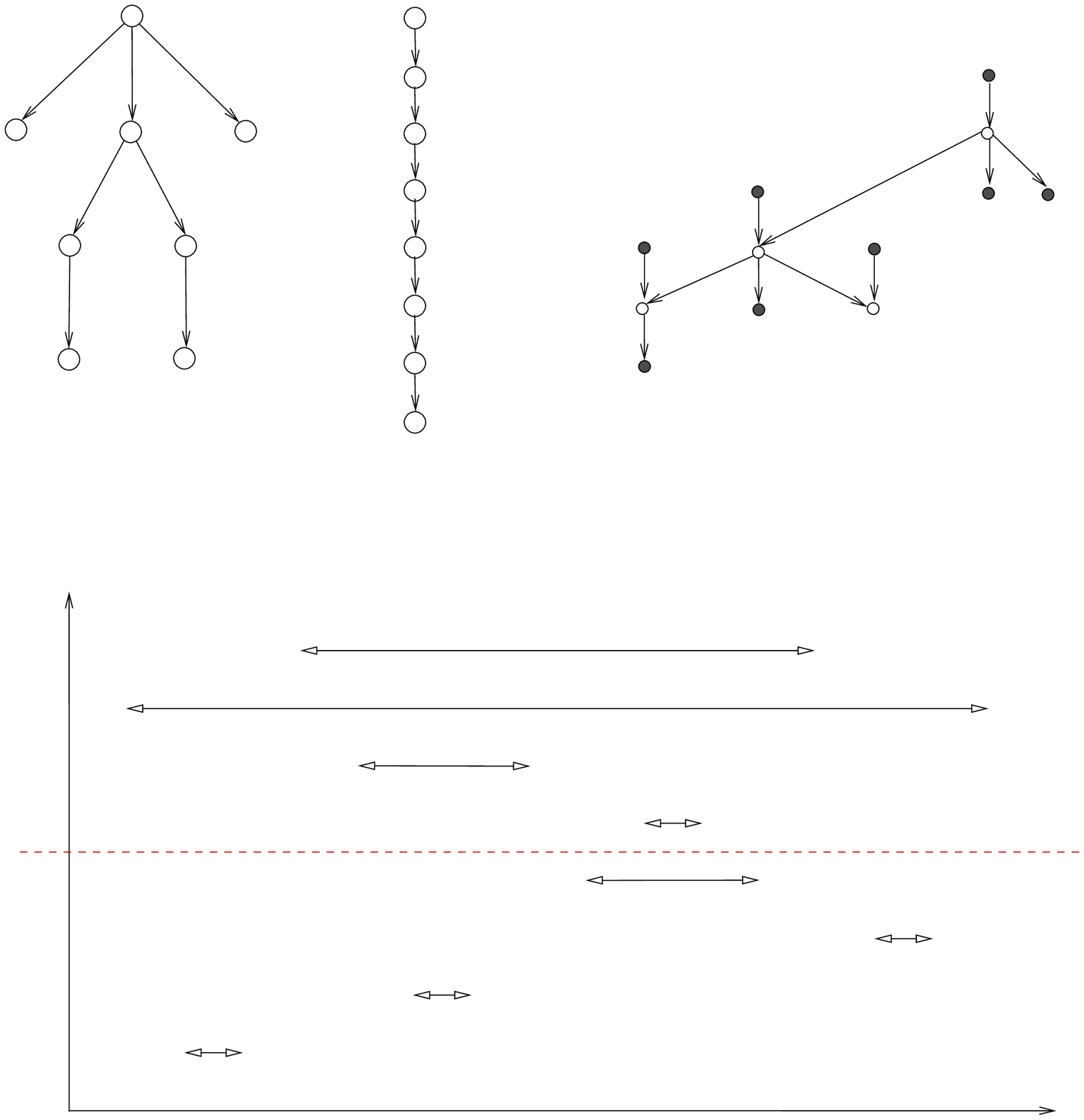}
\end{center}
\caption{The mapping of the vertices of a rooted tree and a dipath to horizontal segments in a 2d rank space and the construction of $\mathcal{J}_{\ell}$.\label{fig:path-tree}}
\end{figure}

As in Section \ref{sec:comb:two-paths} we use a divide-and-conquer method to build $\mathcal{J}$. We will consider $G_1$ to be an out-tree; the in-tree case is handled similarly and yields the same asymptotic bound.
Let $\ell$ be the horizontal line with $x_2$-coordinate equal to $n/2$. A vertex $x$ is \emph{above} $\ell$ if $h(x) \ge n/2$; otherwise ($h(x) < n/2$), $x$ is \emph{below} $\ell$.
We create a subgraph $\mathcal{J}_{\ell}$ of $\mathcal{J}$ that connects the vertices above $\ell$ to the vertices below  $\ell$. To that end, for each vertex $u$  above $\ell$ we create a Steiner vertex $u'$ together with the arc $(u,u')$. Let $z$ be the nearest ancestor of $u$ in $T$ that is above $\ell$. If $z$ exists then we add the arc $(z',u')$. Then, for each vertex $y$ below $\ell$ we locate the nearest ancestor $u$ of $y$ in $T$ that is above $\ell$. If $u$ exists then we add the arc $(u',y)$. See Figure \ref{fig:path-tree}.
Finally, we recurse for the vertices above $\ell$ and for the vertices below $\ell$.

It is not hard to verify the correctness of the above construction. The size of the resulting graph can be bounded by $O(n \log{n})$ as in Section \ref{sec:comb:two-paths}. Furthermore, we can generalize this construction for an unoriented tree and an unoriented path, and accomplish the same $O(n \log{n})$ bound as required by Theorem \ref{thm:combinatorial}(a). (We omit the details which are similar to the more complicated construction of Section \ref{sec:comb:unoriented-trees}.)

\subsection{Two Trees}
\label{sec:comb:two-trees}

The construction of Section \ref{sec:comb:path-tree} can be extended to handle more than one dipath. We show how to apply this extension in order to get an $O(n \log^2{n})$ bound for the join-reachability graph of two rooted trees. We consider the case where $G_1$ is an out-tree and $G_2$ is an in-tree; the other two cases (two out-trees and two in-trees) are handled similarly.

\begin{figure}
\begin{center}
\scalebox{0.7}[0.7]{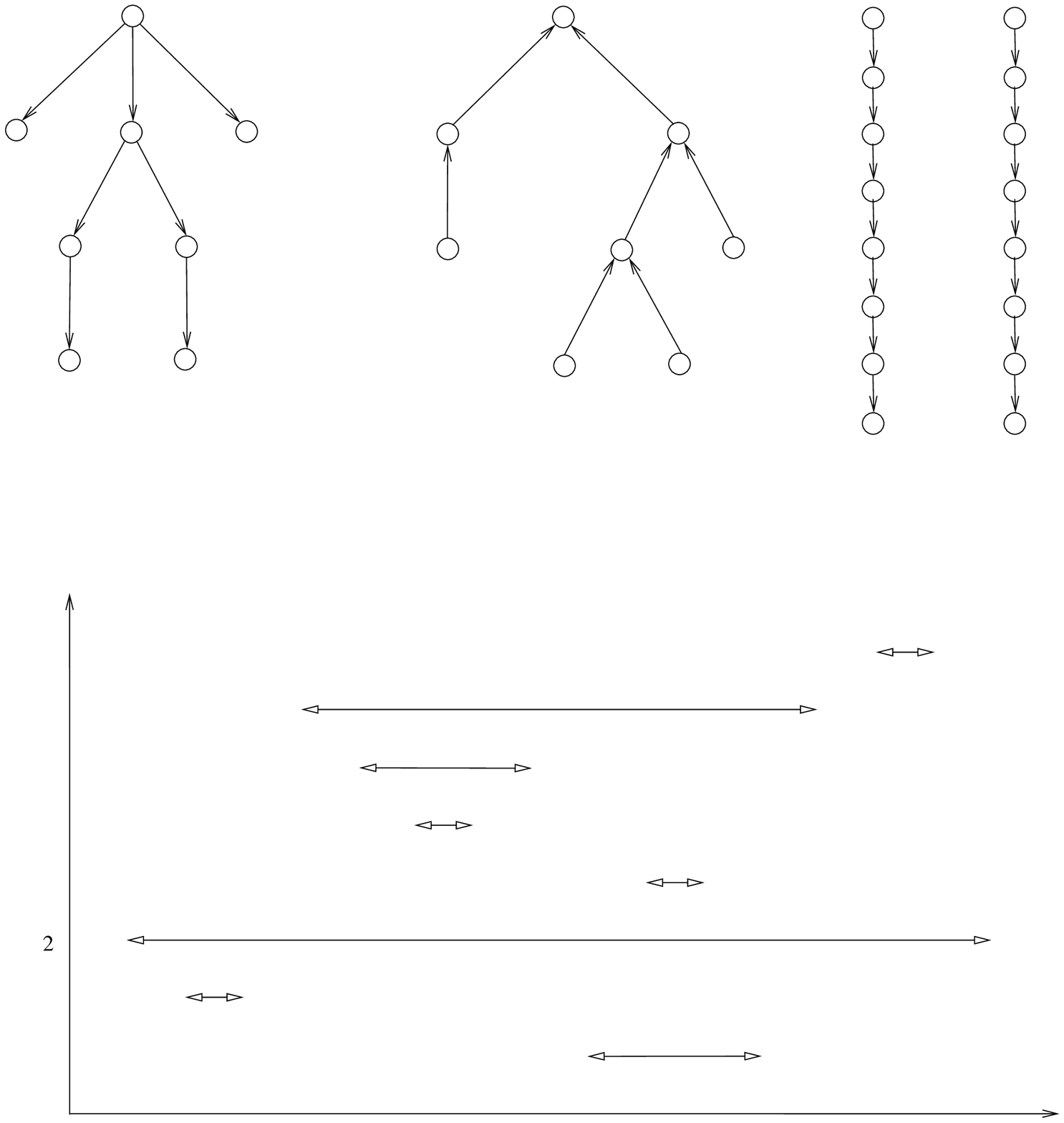}
\end{center}
\caption{The mapping of the vertices of two rooted trees to horizontal segments in a 3d rank space. The value in brackets above the segments correspond to the $x_3$-coordinates.
\label{fig:2-trees-3d}}
\end{figure}

Let $T_1$ and $T_2$ be the corresponding undirected trees. We assign to each vertex $a$ two depth-first search intervals $I_1(a)=[s_1(a),t_1(a)]$ and $I_2(a)=[s_2(a),t_2(a)]$, where $I_j(a)$ corresponds to $T_j$, $j =1,2$.
We create two linear orders (i.e., dipaths), $P_1$ and $P_2$, from the $I_2$-intervals as follows: In $P_1$ the vertices are ordered by decreasing $s_2$-value and in $P_2$ by increasing $t_2$-value. Each vertex $a$ is mapped to an $x_1$-axis-parallel segment $I_1(a) \times x_2(a) \times x_3(a)$ (in three dimensions), where $x_2(a)=h_{P_1}(a)$ and $x_3(a)=h_{P_2}(a)$. Then $a \reach_{\mathcal{J}} b$ if and only if $I_1(b) \subseteq I_1(a)$ and $(x_2(b),x_3(b)) \le (x_2(a),x_3(a))$. See Figure \ref{fig:2-trees-3d}.

Again we employ a divide-and-conquer approach and use the method of Section \ref{sec:comb:path-tree} as a subroutine. The details are as follows. Let $p$ be the plane with $x_3$-coordinate equal to $n/2$. We construct a subgraph $\mathcal{J}_p$ of $\mathcal{J}$ that connects the vertices above $p$ (i.e., vertices $z$ with $x_3(z) \ge n/2$) to the vertices below $p$ (i.e., vertices $z$ with $x_3(z) < n/2$). Then we use recursion for the vertices above $p$ and the vertices below $p$.

\begin{figure}
\begin{center}
\scalebox{0.7}[0.7]{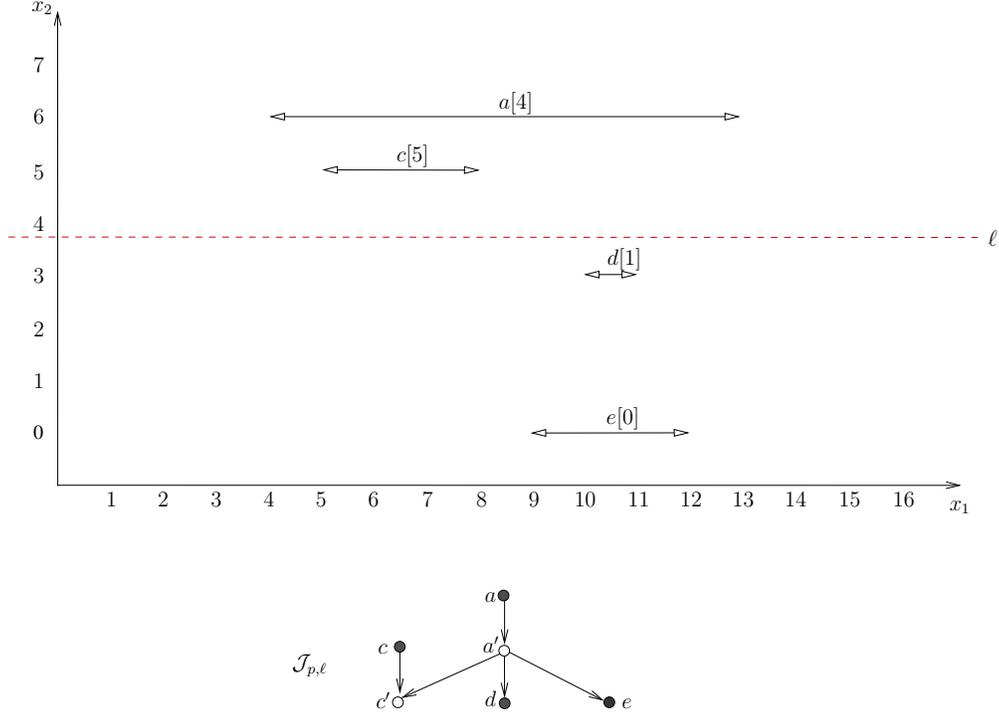}
\end{center}
\caption{The construction of $\mathcal{J}_{p,\ell}$.\label{fig:2-trees-3d-b}}
\end{figure}

We construct $\mathcal{J}_p$ using the method of Section \ref{sec:comb:path-tree} with some modifications. Let $\ell$ be the horizontal line with $x_2$-coordinate equal to $n/2$. We create a subgraph $\mathcal{J}_{p,\ell}$ of $\mathcal{J}_p$ that connects the vertices above $p$ and $\ell$  to the vertices below $p$ and $\ell$. To that end, for each vertex $z$ with $(x_2(z),x_3(z)) \ge (n/2,n/2)$ we create a Steiner vertex $z'$ together with the arc $(z,z')$. Let $u$ be the nearest ancestor of $z$ in $T_1$ such that $(x_2(u),x_3(u)) \ge (n/2,n/2)$. If $u$ exists then we add the arc $(u',z')$. Finally, for each vertex $y$ with $(x_2(y),x_3(y)) < (n/2,n/2)$ we locate the nearest ancestor $z$ of $y$ in $T_1$ such that $(x_2(z),x_3(z)) \ge (n/2,n/2)$. If $z$ exists then we add the arc $(z',y)$. See Figure \ref{fig:2-trees-3d-b}.
Finally, we recurse for the vertices above $\ell$ and for the vertices below $\ell$.

Now we bound the size of our construction. From Section \ref{sec:comb:path-tree} we have that the size of each substructure $\mathcal{J}_p$ is $O(n \log{n})$. Since each vertex participates in $O(\log{n})$ such substructures, the total size is bounded by $O(n\log^2{n})$.

\subsection{Unoriented Trees}
\label{sec:comb:unoriented-trees}

\begin{figure}
\begin{center}
\scalebox{0.7}[0.7]{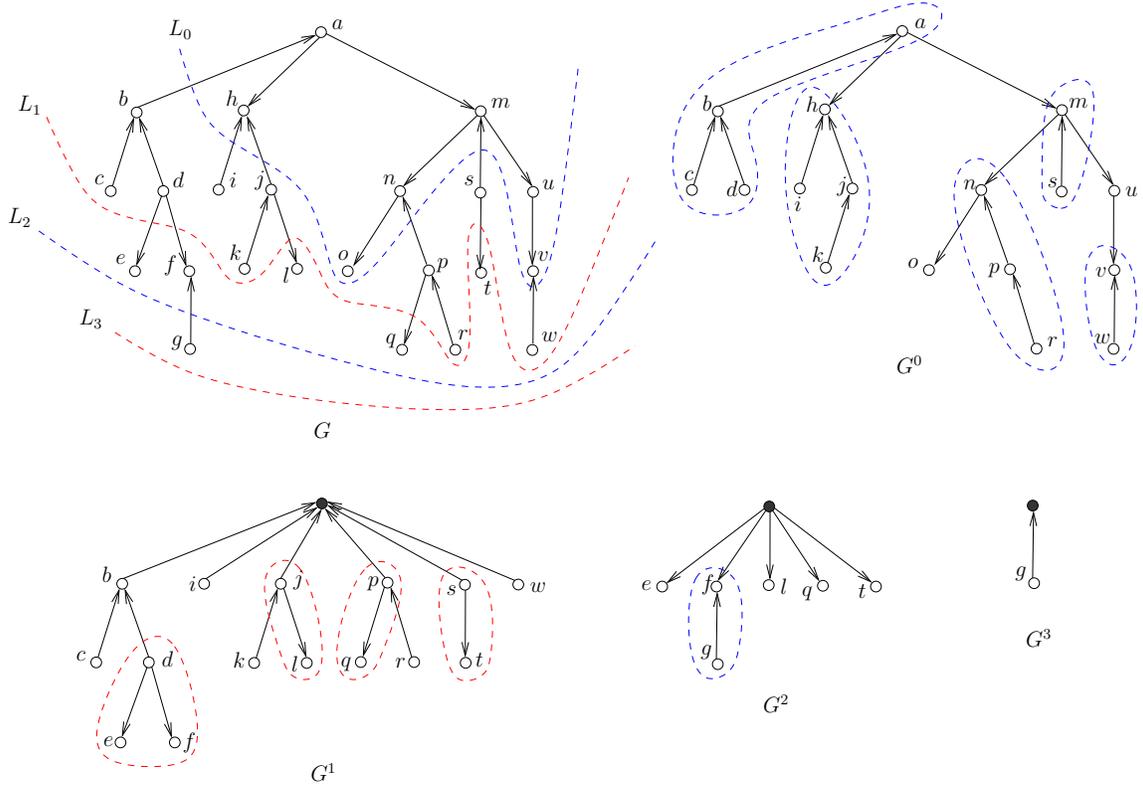}
\end{center}
\caption{An unoriented tree and its sequence of 2-layered tree. Fringe trees are encircled. \label{fig:unoriented-tree}}
\end{figure}

We can reduce the case of unoriented trees to that of rooted trees
by applying Thorup's layer decomposition (see Section \ref{sec:preprocessing}).
We apply this decomposition to both $G_1$ and $G_2$. Let $G_i^0, G_i^2, \ldots, G_i^{\mu_i-1}$ be the sequence of rooted trees produced from $G_i$, $i=1,2$, where each $G_i^j$ is a 2-layered tree. See Figure \ref{fig:unoriented-tree}. For even $j$, $G_i^j$ consists of a \emph{core} out-tree, formed by the arcs directed away from the root, and a collection of \emph{fringe} in-trees. The situation is reversed for odd $j$, where the core tree is an in-tree and the fringe trees are out-trees.
We call a vertex of the core tree a \emph{core vertex}; we call a vertex of a fringe tree (excluding its root) a \emph{fringe vertex}.

We build $\mathcal{J}$ as the union of join-reachability graphs $\mathcal{J}_{i,j}$ for each pair $(G_1^i,G_2^j)$. Each graph $\mathcal{J}_{i,j}$ is constructed similarly to Section \ref{sec:comb:two-trees}, with the exception that we have to take special care for the fringe vertices. (We also remark that in general $\mathcal{J}_{i,j} \neq \mathcal{J}(\{G_1^i,G_2^j\})$.) A vertex $z \in V(G_1^i) \cap V(G_2^j)$ is included in $\mathcal{J}_{i,j}$ if one of the following cases hold: (i) $z$ is a core vertex in at least one of $G_1^i$ and $G_2^j$, or (ii) $z$ is a fringe vertex in both $G_1^i$ and $G_2^j$ and the corresponding fringe trees containing $z$ are either both in-trees or both out-trees. Let $V_{i,j}$ be the vertices in $V(G_1^i) \cap V(G_2^j)$ that satisfy the above condition.

If $V_{i,j} = \emptyset$ then $\mathcal{J}_{i,j}$ is empty.
Now suppose $V_{i,j} \not= \emptyset$. First consider the case where the core of $G_{1}^{i}$ is an out-tree. We contract each fringe in-tree to its root and let the new core supervertex correspond to the vertices of the contracted fringe tree. Let $\hat{G}_{1}^{i}$ be the out-tree produced from this process. Equivalently, if the core of $G_{1}^{i}$ is an in-tree then the contraction of the fringe out-trees produces an in-tree $\hat{G}_{1}^{i}$. We repeat the same process for $G_{2}^{j}$. Next, we assign a depth-first search interval $I_1(z)$ to each vertex $z$ in $\hat{G}_{1}^{i}$ and a depth-first search interval $I_2(z)$ to each vertex $z$ in $\hat{G}_{2}^{j}$, as in Section \ref{sec:comb:two-trees}. The vertices in $V_{i,j}$ are assigned a depth-first search interval in both trees, and therefore can be mapped to horizontal segments in a 3d space, as in Section \ref{sec:comb:two-trees}. Hence, we can employ the method of Section \ref{sec:comb:two-trees} with some necessary changes that involve the fringe vertices. Let $z \in V_{i,j}$ be a fringe vertex in at least one of $G_{1}^{i}$ and $G_{2}^{j}$. If the fringe tree containing $z$ is an in-tree then we only include in $\mathcal{J}_{i,j}$ arcs leaving $z$; otherwise we only include arcs entering $z$.

Finally we need to show that the size of the resulting graph is $O(n \log^{2}{n})$. This follows from the fact that each subgraph $\mathcal{J}_{i,j}$ has size $O(n \log^{2}{n})$ and that each vertex can appear in at most four such subgraphs. Theorem \ref{thm:combinatorial}(b) follows.

\subsection{Planar Digraphs}
\label{sec:comb:planar-graphs}

Now we turn to planar digraphs and combine our previous constructions with Thorup's reachability oracle~\cite{planar-reach:thorup}. From this combination we derive the bounds stated in Theorem \ref{thm:combinatorial}(c) and (d). First we need to provide some details for the reachability oracle of \cite{planar-reach:thorup}.

Let $G$ be a planar digraph, and let $G^0, G^1, \ldots, G^{\mu-1}$ be the sequence of 2-layered digraphs produced from $G$ as described in Section \ref{sec:preprocessing}.
Consider one of these digraphs $G^{i}$. The next step is to obtain a separator decomposition of $G^{i}$. To that end, we treat $G^{i}$ as an undirected graph and compute a separator $S$ whose removal separates $G^{i}$ into components, each with at most half the vertices. The separator $S$ consists of three root paths of a spanning tree of $G^{i}$ rooted at $r_0$. Because $G^{i}$ is 2-layered, each root path in $S$ corresponds to at most two dipaths in $G^{i}$. The key idea now is to process each separator dipath $Q$ and find the connections between $V(G^{i})$ and $Q$. \ignore{To that end, we identify the vertices of $Q$ by their rank in $Q$.} For each $v \in V(G^{i})$ two quantities are computed: (i) $\mathrm{from}_v[Q]$ which is equal to $r_Q(u)$, where $u \in Q$ is the vertex with the highest rank in $Q$ such that $u \reach_{G^{i}} v$, and (ii) $\mathrm{to}_v[Q]$ which is equal to $r_Q(u)$, where $u \in Q$ is the vertex with the lowest rank in $Q$ such that $v \reach_{G^{i}} u$. Clearly there is a path from $a$ to $b$ that passes though $Q$ if and only if $\mathrm{to}_a[Q] \le \mathrm{from}_b[Q]$. The same process is carried out recursively for each component of $G^{i} \setminus V(S)$. The depth of this recursion is $O(\log n)$, so each vertex is connected to $O(\log n)$ separator dipaths. The space and construction time for this structure is $O(n \log n)$.

Now we consider how to construct a join-reachability graph when $G_1$ is a planar digraph. We begin with the case where $G_2$ is a dipath.
First we perform the layer decomposition of $G_1$ and construct the corresponding graph sequence $G^{0}_{1},G^{1}_{1},\ldots,G^{\mu-1}_{1}$. Then we form pairs of digraphs $P_i= \{ G^{i}_{1},G^{i}_{2} \}$ where $G^{i}_{2}$ is a dipath containing only the vertices in $V(G^{i}_{1})$ in the order they appear in $G_2$. Clearly $a \reach_{\mathcal{J}} b$ if and only if $a \reach_{\mathcal{J}_{\iota(b)-1}} b$ or $a \reach_{\mathcal{J}_{\iota(b)}} b$, where $\mathcal{J}_i$ is the join-reachability graph of $P_i$. Then $\mathcal{J}$ is formed from the union of $\mathcal{J}_0,\ldots,\mathcal{J}_{\mu-1}$.

To construct $\mathcal{J}_i$ we perform the separator decomposition of $G^{i}_{1}$, so that each vertex is associated with $O(\log n)$ separator dipaths. Let $Q$ be such a separator dipath. Also, let $V_Q$ be the set of vertices that have a successor or a predecessor in $Q$.
We build a subgraph $\mathcal{J}_{i,Q}$ of $\mathcal{J}_i$ for the vertices in $V_Q$; $\mathcal{J}_i$ is formed from the union of the subgraphs $\mathcal{J}_{i,Q}$ for all the separator dipaths of $G^{i}_{1}$. The construction of $\mathcal{J}_{i,Q}$ is carried out
as follows. Let $z \in V_Q$. If $z$ has a predecessor in $Q$ then we create a vertex $z^{-}$ which is assigned coordinates $x_1(z^{-})=\mathrm{from}_{z}[Q]$ and $x_2(z^{-}) = r_{G_2}(z)$, and add the arc $(z,z^{-})$. Similarly, if $z$ has a successor in $Q$ then we create a vertex $z^{+}$ which is assigned coordinates $x_1(z^{+})=\mathrm{to}_{z}[Q]$ and $x_2(z^{+}) = r_{G_2}(z)$, and add the arc $(z^{+},z)$.

Now we can use the method of Section \ref{sec:comb:two-paths} to build the rest of $\mathcal{J}_{i,Q}$, so that $a \reach_{\mathcal{J}_{i,Q}} b$ if and only if $(x_1(a^{+}),x_2(a^{+})) \le (x_1(b^{-}),x_2(b^{-}))$.
Let $\ell$ be the vertical line with $x_1$-coordinate equal to $n/2$. The first step is to construct the subgraph of $\mathcal{J}_{i,Q}$ that connects the vertices $a^{+}$ with $x_1(a^{+}) \le n/2$ to the vertices $b^{-}$ with $x_1(b^{-}) \ge n/2$.
For each such $b^{-}$ we create a Steiner vertex $b'$ and add the arc $(b',b^{-})$. Also, we assign to $b'$ the coordinates $(n/2,x_2(b^{-}))$.
We connect these Steiner vertices in a dipath starting from the vertex with the lowest $x_2$-coordinate. Next, for each vertex $a^{+}$ with $x_1(a^{+}) \le n/2$ we locate the Steiner vertex $b'$ with the smallest $x_2$-coordinate such that $x_2(a^{+}) \le x_2(b')$. If $b'$ exists we add the arc $(a^{+},b')$.
Finally we recurse for the vertices with $x_1$-coordinate in $[1,n/2)$ and for the vertices with $x_1$-coordinate in $(n/2,n]$.

It remains to bound the size of $\mathcal{J}$. From Section \ref{sec:comb:two-paths}, we have $|\mathcal{J}_{i,Q}| = O(|V_Q| \log |V_Q|)$. Moreover, the bound $\sum_Q|V_Q| = O(|V(G^{i}_1)| \log |V(G^{i}_1)|)$, where the sum is taken over all separator paths of $G^{i}_1$, implies $|\mathcal{J}_{i}| \le \sum_Q |\mathcal{J}_{i,Q}| = O(|V(G^{i}_1)| \log^2 |V(G^{i}_1)|)$. Finally, since $\sum_i |V(G^{i}_1)| = O(n)$ we obtain $|\mathcal{J}| \le \sum_i |\mathcal{J}_{i}| = O(n \log^2 n)$.

We handle the case where $G_2$ is an unordered dipath as noted in Section \ref{sec:comb:two-paths}, which implies Theorem \ref{thm:combinatorial}(c).
The methods we developed here in combination with the structures of Section \ref{sec:comb:unoriented-trees} result to a join-reachability graph of size $O(n \log^3 n)$ for a planar digraph and an unoriented tree. The same bound of $O(n \log^3 n)$ is achieved for two planar digraphs, as stated in Theorem \ref{thm:combinatorial}(d).

\subsection{General Graphs}
\label{sec:comb:general-graphs}

A technique that is used to speed up transitive closure and reachability computations is to cover a digraph with simple structures such as dipaths, chains, or trees (e.g., see \cite{transitive-db:abj}). Such techniques are well-suited to our framework as they can be combined with the structures we developed earlier. We also remark that the use of the preprocessing steps of Section \ref{sec:preprocessing} reduces the problem from general digraphs to acyclic and 2-layered digraphs.
In this section we describe how to obtain join-reachability graphs with the use of dipath covers. This gives the bounds stated in Theorem \ref{thm:combinatorial}(e)-(g); similar results can be derived with the use of tree covers. Again for simplicity, we first consider the case where $G_1$ is a general digraph and $G_2$ is a dipath.

A \emph{dipath cover} is a decomposition of a digraph into vertex-disjoint dipaths. Let $P^{1}_1, P^{2}_{1}, \ldots P^{\kappa_1}_{1}$ be a dipath cover of $G_1$. For each vertex $v$ and each path $P^{i}_{1}$ we compute $\mathrm{from}_v[P^{i}_{1}]$, i.e., $r_{P^{i}_{1}}(z)$ where $z \in P^{i}_{1}$ is the vertex with the highest rank in $P^{i}_{1}$ such that $z \reach_{G_1} v$.
Let $P^{i}_{2}$ be the dipath that consists of the vertices in $P^{i}_{1}$ ordered by increasing rank in $G_2$. Also, set $\mathrm{from}_v[P^{i}_{2}] = r_{P^{i}_{2}}(z)$ where $z \in P^{i}_{2}$ is the vertex with the largest rank such that $r_{G_{2}}(z) \le r_{G_{2}}(v)$.
Let $V_{P^{i}_{1}}$ be set of vertices that have a predecessor in $P^{i}_{1}$.
We build a subgraph $\mathcal{J}_{i}$ of $\mathcal{J}$ that connects the vertices of $P^{i}_{1}$ to $V_{P^{i}_{1}}$. Then $\mathcal{J}$ is formed from the union of the subgraphs $\mathcal{J}_{i}$.
For each $z \in V_{P^{i}_{1}}$ we create a vertex $z^{-}$ which is assigned coordinates $x_1(z^{-})= \mathrm{from}_{z}[P^{i}_{1}]$ and $x_2(z^{-}) = \mathrm{from}_{z}[P^{i}_{2}]$, and add the arc $(z^{-},z)$. Also, for each $z \in P^{i}_1$ we create a vertex $z^{+}$ which is assigned coordinates $x_1(z^{+})=r_{P^{i}_{1}}(z)$ and $x_2(z^{+})=r_{P^{i}_{2}}(z)$, and add the arc $(z,z^{+})$. Now we can build a join-reachability graph, so that $a \reach_{\mathcal{J}_{i}} b$ if and only if $(x_1(a^{+}),x_2(a^{+})) \le (x_1(b^{-}),x_2(b^{-}))$, as in Section \ref{sec:comb:planar-graphs}.

The size of this graph is bounded by $\sum_{i} |V_{P^{i}_{1}}| \log |V_{P^{i}_{1}}| = O(\kappa_1 n \log n)$, which implies the result of Theorem \ref{thm:combinatorial}(e). We can extend this method to handle two general digraphs and obtain the bound of Theorem \ref{thm:combinatorial}(g). The case where $G_2$ is planar digraph is handled by combining the above method with the techniques of Section \ref{sec:comb:planar-graphs}, resulting to Theorem \ref{thm:combinatorial}(f).

\section{Data Structures for Join-Reachability}
\label{sec:ds}

Now we deal with the data structure version of the join-reachability problem. Our goal is to construct an efficient data structure for $\mathcal{J} \equiv \mathcal{J}(\{G_1,G_2\})$ such that given a query vertex $b$ it can report all vertices $a$ satisfying $a \reach_{\mathcal{J}} b$.
We state the efficiency of a structure using the notation $\langle s(n),q(n,k) \rangle$ which refers to a data structure with $O(s(n))$ space and $O(q(n,k))$ query time for reporting $k$ elements. In order to design efficient join-reachability data structures we apply the techniques we developed in Section \ref{sec:combinatorial}. The bounds that we achieve this way are summarized in the following theorem.

\begin{theorem}
\label{thm:ds}
Given two digraphs $G_1$ and $G_2$ with $n$ vertices we can construct join-reachability data structures with the following efficiency:
\begin{itemize}
\ignore{\item[(a)] $\langle n, k \rangle$ when $G_1$ is an unoriented dipath or an out-tree and $G_2$ is an unoriented dipath.\ignore{\vspace{-2ex}}}
\item[(a)] $\langle n, k \rangle$ when $G_1$ is an unoriented tree and $G_2$ is an unoriented dipath.\ignore{\vspace{-2ex}}
\item[(b)] $\langle n, \log n + k \rangle$ when $G_1$ is an out-tree and $G_2$ is an unoriented tree.\ignore{\vspace{-2ex}}
\item[(c)] $\langle n \log^{\varepsilon}{n},  \log{\log{n}} + k \rangle$ (for any constant $\varepsilon>0$), when $G_1$ and $G_2$ are unoriented trees.\ignore{\vspace{-2ex}}
\item[(d)] $\langle n \log n, k \log{n} \rangle$ when $G_1$ is planar digraph and $G_2$ is an unoriented tree.\ignore{\vspace{-2ex}}
\ignore{\item[(e)] $\langle n \log n, k \log{n} + \log^2{n} \rangle$ when $G_1$ is planar digraph and $G_2$ is an unoriented tree.\ignore{\vspace{-2ex}}}
\item[(e)] $\langle n \log^2 n, k \log^2{n} \rangle$ when both $G_1$ and $G_2$ are planar digraphs.\ignore{\vspace{-2ex}}
\item[(f)] $\langle n \kappa_1, k \rangle$ when $G_1$ is a general digraph that can be covered with $\kappa_1$ vertex-disjoint dipaths and $G_2$ is an unoriented tree.\ignore{\vspace{-2ex}}
\item[(g)] $\langle n (\kappa_1 + \log n), k\kappa_1\log{n} \rangle$ or $\langle n \kappa_1 \log n, k \log n \rangle$ when $G_1$ is a general digraph that can be covered with $\kappa_1$ vertex-disjoint dipaths and $G_2$ is planar digraph.\ignore{\vspace{-2ex}}
\item[(h)] $\langle n (\kappa_1+\kappa_2), \kappa_1 \kappa_2 + k \rangle$ or $\langle n \kappa_1 \kappa_2, k \rangle$ when each $G_i$, $i=1,2$, is a digraph that can be covered with $\kappa_i$ vertex-disjoint dipaths.
\end{itemize}
\end{theorem}

Next we provide the constructions that prove the bounds stated in Theorem \ref{thm:ds}. Throughout this section $k$ denotes the size of the output of a join-reachability reporting query.

\subsection{Two Paths}
\label{sec:ds:paths}

Let $G_1$ and $G_2$ be two dipaths.
We use the mapping of Section \ref{sec:complexity}. Recall that each vertex $a$ is mapped to a point $(x_1(a),x_2(a))$ on an $n \times n$ grid so that $a \reach_\mathcal{J} b$ if and only if $(x_1(a),x_2(a)) \le (x_1(b),x_2(b))$. This is a \emph{two-dimensional point dominance problem} that can be solved optimally with a Cartesian tree~\cite{scaling:gbt84}. Thus, we immediately get an $\langle n, k \rangle$ join-reachability structure for two dipaths.
We provide the details of this structure as we will need them in later constructions. A Cartesian tree $T$ is a binary tree defined recursively as follows. The root of $T$ is the point $a$ with minimum $x_2$-coordinate. The left subtree of the root is a Cartesian tree for the points $b$ with $x_1(b)<x_1(a)$ and the right subtree of the root is a Cartesian tree for the points $b$ with $x_1(b)>x_1(a)$. Clearly this structure uses linear space, and moreover it can be constructed in linear time~\cite{scaling:gbt84}.
The reporting algorithm uses the following property of Cartesian trees. Consider two points $a$ and $b$, and let $c$ be the point with minimum $x_2$-coordinate such that $x_1(a) \le x_1(c) \le x_1(b)$. Then, $c=\nca_T(a,b)$. Now let $\zeta$ be the point with the smallest $x_1$-coordinate. In order to find all points $a$ such that $(x_1(a),x_2(a)) \le (x_1(b),x_2(b))$ we first locate $y=\nca_T(\zeta,b)$. The returned point $y$ has the smallest $x_2$-coordinate in the $x_1$-range $[0,x_1(b)]$. If $x_2(y) > x_2(b)$ then the answer is null and we stop our search. Otherwise we return $y$ and search recursively in the $x_1$-ranges $[0,x_1(y)-1]$ and $[x_1(y)+1,x_1(b)]$. Using the fact that nearest common ancestor queries in a tree can be answered in constant time after linear time preprocessing~\cite{nca:ht}, it follows that the time to report $k$ vertices is $O(k)$.

As in Section \ref{sec:comb:two-paths}, we can achieve the same bounds when $G_1$ and $G_2$ are unoriented dipaths by splitting them into maximal subpaths consisting of arcs with the same orientation. \ignore{Hence, the first case of Theorem \ref{thm:ds}(a) follows.}

\subsection{Tree and Path}
\label{sec:ds:path-tree}

Next we consider the case where $G_1$ is a rooted tree and $G_2$ is a dipath. As in Section \ref{sec:comb:path-tree}, we note that a rooted tree can be described by two linear orders, and therefore we can get an $\langle n , \log{n}+k \rangle$ solution using a three-dimensional dominance reporting structure~\cite{multidim-dom:jms04}. Here we develop an alternative method that reduces the dimension of our problem and as a result it achieves an
$\langle n , k \rangle$ bound. Furthermore, this method can be extended to give more efficient structures for two trees (compared to four-dimensional dominance reporting~\cite{multidim-dom:jms04}). We will distinguish two cases depending on whether $G_1$ is an out-tree or an in-tree. In any case, let $T$ be the rooted tree that results from $G_1$ after removing arc directions. We associate each vertex $x \in T$ with a label $h(x)=h_{G_2}(x)$, the height of $x$ in $G_2$. For an in-tree we wish to support the following query: Given a vertex $b$ and a label $j$ find all vertices $a \in T(b)$ with $h(a)>j$. Equivalently, for an out-tree the query algorithm needs to find all ancestors $a$ of $b$ in $T$ with $h(a)>j$. We present a geometry-based method, which achieves $O(\log{n}+k)$ reporting time for an in-tree and $O(k)$ for an out-tree. An alternative method, based on a heavy-path decomposition of $T$~\cite{dyntrees:st83}, is given in Appendix \ref{appendix:path-decomposition}.

We use the mapping of Section \ref{sec:comb:path-tree}. Each vertex $a$ is assigned a depth-first search interval $I(a)=[s(a),t(a)]$ in $T$ and is mapped to the $x_1$-axis-parallel segment $S(a)=I(a) \times h(a)$. Now the choice of the structure we use depends on the arc directions in $G_1$.
For an out-tree we have that $a \rightsquigarrow_{\mathcal{J}} b$ if and only if $S(a)$ is above $S(b)$ and the $x_1$-projection of $S(a)$ covers the $x_1$-projection of $S(b)$. The fact that interval endpoints are distinct implies that
$a \rightsquigarrow_{\mathcal{J}} b$ if and only if the vertical ray $v_b$ emanating from $(s(b),h(b))$ towards the $(+x_2)$-direction intersects $S(a)$.
Indeed, if $a \rightsquigarrow_{\mathcal{J}} b$ then $h(b) \le h(a)$ and $b \in T(a)$, so $I(b) \subseteq I(a)$. Similarly, if $S(a)$ is above $S(b)$ and $I(b) \subseteq I(a)$ then $v_b$ intersects $S(a)$. Therefore, we have reduced our problem to a \emph{planar segment intersection} problem. We can get an
$\langle n,  k \rangle$ structure by adapting either the hive graph of Chazelle~\cite{filt:c86} or the persistence-based planar point location structure of Sarnak and Tarjan~\cite{persistent:st}. Both these data structures require $O(n\log n)$ preprocessing time  as they need to sort the endpoint coordinates. In our case sorting is not necessary, since the $x_1$-coordinates are produced in sorted order by the depth-first search, and the $x_2$-coordinates correspond to the height of the vertices in $G_2$. Hence our preprocessing time is $O(n)$. Furthermore, the reporting time using either the hive graph or the persistence-based structure is $O(\log{n}+k)$, where the $\log{n}$ term is due to a point location query. In our case this term can be reduced to constant; point location is not necessary since the segment endpoints are the only possible query locations. Hence our reporting time is $O(k)$. \ignore{This implies Theorem \ref{thm:ds}(a).}

We turn to the case where $G_1$ is an in-tree. Here we have that $a \rightsquigarrow_{\mathcal{J}} b$ if and only if $S(a)$ is below $S(b)$ and the $x_1$-projection of $S(b)$ covers the $x_1$-projection of $S(a)$. Since the interval endpoints are distinct we have
$a \rightsquigarrow_{\mathcal{J}} b$ if and only if the endpoints of $S(a)$ are contained inside the rectangle $[s(b),t(b)] \times [0,h(b)]$.
This is a \emph{two-dimensional grounded range search} problem (one side of the query rectangle always lies on the $x_1$-axis). Since we have integer coordinates in $[1,2n] \times [0,n-1]$ we can get an $\langle n, k \rangle$ structure again with the use of a Cartesian tree~\cite{scaling:gbt84}.

The $\langle n,  k \rangle$ bound is also achieved when $G_1$ is an unoriented tree, as stated in Theorem \ref{thm:ds}(a), by applying the method of Section \ref{sec:comb:unoriented-trees}.
Let $G_1^0, G_1^2, \ldots, G_1^{\mu_i-1}$ be the sequence of 2-layered rooted trees produced from $G_1$. We construct a join-reachability structure for each pair $P_i = \{ G_{1}^{i}, G_{2}^{i} \}$, where $G_{2}^{i}$ is a dipath containing only the vertices in $V(G_1^{i})$ in the order they appear in $G_2$. A query for a vertex $b$ needs to search the structures for the pairs $P_{\iota(b)-1}$ and $P_{\iota(b)}$.
The structure for $P_i$ is constructed as follows. We contract each fringe tree to its root and let the new core supervertex correspond to the vertices of the contracted fringe tree. Let $\hat{G}_{1}^{i}$ be the tree produced from this process. Next, we assign a depth-first search interval $I_1(z)$ to each vertex $z$ in $\hat{G}_{1}^{i}$, and map $z$ to the $x_1$-axis-parallel segment $I_1(z) \times x_2(z)$, where $x_2(z)=h_{G_{2}^{i}}(z)$. Using this mapping we can construct the data structures developed above depending on whether $\hat{G}_1^{i}$ is an out-tree or an in-tree. One important detail is that if $\hat{G}_1^{i}$ is an in-tree then the data structure for $P_i$ does not store the segments that correspond to fringe vertices; the segment of such a fringe vertex $z$ is needed however in order to answer a join-reachability query for $z$.
Equivalently, if $\hat{G}_1^{i}$ is an out-tree and the query vertex $b$ is an fringe in-tree of $G_1^{i}$ then we do not search the structure for $P_i$.

\subsection{Two Trees}
\label{sec:ds:two-trees}

We extend the method of Section \ref{sec:ds:path-tree} in order to deal with two rooted trees $G_1$ and $G_2$. We distinguish three cases depending on the type, in-tree or out-tree, of each tree. Then, by applying the layer decomposition method of Section \ref{sec:comb:unoriented-trees}, we can extend our structures to handle unoriented trees. This way we achieve the bounds
stated in Theorem \ref{thm:ds}(b) and (c).

\begin{figure}
\begin{center}
\scalebox{0.7}[0.7]{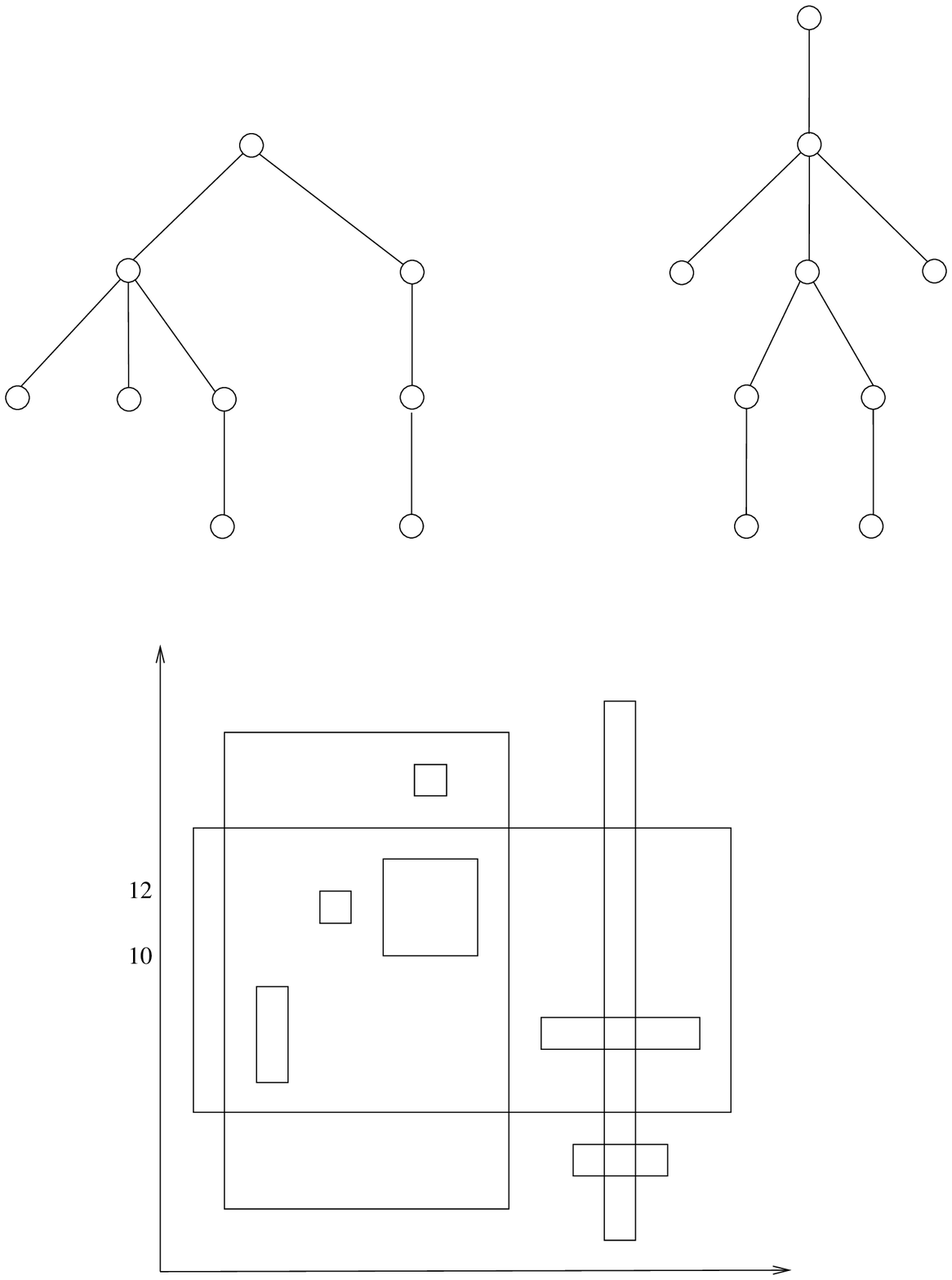}
\end{center}
\caption{Example of the mapping of Section \ref{sec:ds:two-trees}
\label{fig:two-trees}}
\end{figure}

Let $T_1$ and $T_2$ be the corresponding undirected trees. We assign each vertex $a$ two depth-first search intervals $I_1(a)=[s_1(a),t_1(a)]$ and $I_2(a)=[s_2(a),t_2(a)]$, where $I_j(a)$ corresponds to $T_j$, for $j = 1,2$.
We use the two intervals $I_1(a)=[s_1(a),t_1(a)]$ and $I_2(a)=[s_2(a),t_2(a)]$ to map each vertex $a$ to an axis-parallel rectangle $R(a)=I_1(a) \times I_2(a)$. See Figure \ref{fig:two-trees}.
Again we exploit the fact that for any two vertices $a$ and $b$, the intervals $I_j(a)$ and $I_j(b)$ are either disjoint or one contains the other.
If $I_1(a) \cap I_1(b) = \emptyset$ or $I_2(a) \cap I_2(b) = \emptyset$, then $R(a)$ and $R(b)$ do not intersect.
Now suppose that both $I_1(a) \cap I_1(b) \not= \emptyset$ and $I_2(a) \cap I_2(b) \not= \emptyset$.
Without loss of generality, consider that $I_1(b) \subseteq I_1(a)$. If $I_2(b) \subseteq I_2(a)$ then $R(b)$ is contained in $R(a)$. Otherwise, if $I_2(a) \subseteq I_2(b)$ then both horizonal edges of $R(a)$ intersect both vertical edges of $R(b)$. Next, we distinguish three cases depending on the type of the two trees.

First suppose that both $G_1$ and $G_2$ are out-trees. Then
$a \rightsquigarrow_{\mathcal{J}} b$ implies $b \in T_1(a)$ and $b \in T_2(a)$. So here we have  $I_1(b) \subseteq I_1(a)$ and $I_2(b) \subseteq I_2(a)$, thus $R(b)$ is contained in $R(a)$. In particular, the rectangle arrangement has the property that
$a \rightsquigarrow_{\mathcal{J}} b$ if and only if $R(a)$ encloses a corner of $R(b)$.
This property implies that we have a \emph{two-dimensional point enclosure}: In order to report
all vertices $a$ such that $a \rightsquigarrow_{\mathcal{J}} b$ we need to find all rectangles $R(a)$ that enclose a corner of $R(b)$. To that end, we can use the point enclosure structure of Chazelle~\cite{filt:c86} to get an $\langle n,  \log{n} + k \rangle$ join-reachability structure.

Next, suppose that $G_1$ is an out-tree and $G_2$ is an in-tree. In this case $a \reach_{\mathcal{J}} b$ if and only if $b \in T_1(a)$ and $a \in T_2(b)$, which implies $I_1(b) \subseteq I_1(a)$ and $I_2(a) \subseteq I_2(a)$. Thus, $R(a)$ intersects $R(b)$. Furthermore, the properties of the depth-first search intervals imply that $a \rightsquigarrow_{\mathcal{J}} b$ if and only if the segment $s_1(b) \times I_2(b)$ intersects $I_1(a) \times s_2(a)$.
This is an orthogonal segment intersection problem, for which we can get an $\langle n,  k \rangle$ join-reachability structure as in Section \ref{sec:ds:path-tree}.

The last case is when $G_1$ and $G_2$ are in-trees. Now $a \reach_{\mathcal{J}} b$ if and only if $a \in T_1(b)$ and $a \in T_2(b)$. Then we have $I_1(a) \subseteq I_1(b)$ and $I_2(a) \subseteq I_2(b)$, which implies that $a \rightsquigarrow_{\mathcal{J}} b$ if and only if $R(b)$ encloses a corner of $R(a)$. Thus, our reporting query reduces to \emph{orthogonal range searching}. Here the results of Alstrup et al.~\cite{range-searching:abr} imply an $\langle n \log^{\varepsilon}{n},  \log{\log{n}} + k \rangle$ join-reachability structure (for any constant $\varepsilon>0$).

\subsection{Planar Digraphs}
\label{sec:ds:planar}

With the help of Thorup's reachability oracle~\cite{planar-reach:thorup} we can develop efficient structures for join-reachability in planar digraphs. Suppose first that $G_2$ is a dipath.
We perform the layer decomposition of $G_1$ and construct the corresponding graph sequence $G_{1}^{0},G_{1}^{1},\ldots,G_{1}^{\mu-1}$. Then we form pairs of digraphs $P_i= \{ G_{1}^{i},G_{2}^{i} \}$ where $G_{2}^{i}$ is a dipath containing only the vertices in $V(G_{1}^{i})$ in the order they appear in $G_2$. Clearly $a \reach_{\mathcal{J}} b$ if and only if $a \reach_{\mathcal{J}_{\iota(b)-1}} b$ or $a \reach_{\mathcal{J}_{\iota(b)}} b$, where $\mathcal{J}_i$ is the join-reachability graph of $P_i$.
For each pair $P_i$ we build a join-reachability structure. In order to answer a reporting query for $b$ we query the structures for $P_{\iota(b)-1}$ and $P_{\iota(b)}$ independently and return the union of the results. It remains to describe the structure for a pair $P_i=\{ G_{1}^{i},G_{2}^{i} \}$. We perform the separator decomposition of $G_{1}^{i}$, so that each vertex is associated with $O(\log n)$ separator dipaths. For each vertex $v \in V(G_{1}^{i})$ we record a set $S(v)$ containing the separator dipaths $Q$ that reach $v$ together with the number $\mathrm{from}_v[Q]$ (see Section \ref{sec:comb:planar-graphs}).
For each separator dipath $Q$ we record the vertices $v$ that reach $Q$ together with the numbers $\mathrm{to}_v[Q]$.
Next, for each separator dipath $Q$ we build the data structure of Section ~\ref{sec:ds:paths} for the vertices that reach $Q$. Each such vertex $a$ receives coordinates $(x_1(a),x_2(a))$ where $x_1(a) = \mathrm{to}_a[Q]$ and $x_2(a)$ is the rank of $a$ in $G_2$ among the vertices that reach $Q$. Now we can report the vertices that reach $b$ through $Q$ by finding the vertices $a$ that satisfy $(x_1(a),x_2(a)) \le (\mathrm{from}_b[Q],x_2(b))$. To that end, we use a Cartesian tree $T$ as in Section~\ref{sec:ds:paths}. Here we need to modify this structure in order to allow points with identical $x_1$-coordinates. Since the $x_1$-coordinates are integers in the range $[0,|Q|-1]$ we find for each integer $i$ in that range the point $a_i$ with $x_1(a_i)=i$ and minimum $x_2$-coordinate. Then we build a Cartesian tree for the points $a_i$, $0\le i \le |Q|-1$. Also, we associate with $a_i$ a list of the remaining points with $x_1$-coordinate equal to $i$ in increasing $x_2$-coordinate. Next, in order to initiate the search we also need to locate the vertex $c$ with $x_1(c)=\mathrm{from}_b[Q]$. We can do that easily in $O(1)$ time by using an array of size $|Q|$ to map the $x_1$-coordinates to the corresponding locations in $T$. Recall that the basic step of the reporting algorithm is to locate the point with smallest $x_2$-coordinates in an $x_1$-range $[\alpha,\beta]$. If $y$ is the corresponding point, then we check if $x_2(y) \le \mathrm{from}_b[Q]$. If this is the case, then we report $y$ and search the list associated with $y$ and report all points with $x_2(z) \le \mathrm{from}_b[Q]$. Clearly the reporting time for $k$ points is still $O(k)$. Also the required space and preprocessing is $O(|V(G_{1}^{i})|)$. Therefore, the asymptotic preprocessing time and space are the same as in Thorup's structure, i.e., $O(n \log n)$. Finally we need to specify how to report all vertices $a$ such that $a \reach_{\mathcal{J}} b$. We query the structures for $P_{\iota(b)-1}$ and $P_{\iota(b)}$. To perform a query for $P_i$ we use the list of separator dipaths that reach $b$, and for each such dipath $Q$ we use the corresponding Cartesian tree to report the vertices $a$ that satisfy $(x_1(a),x_2(a)) \le (\mathrm{from}_b[Q],x_2(b))$. Let $k_Q$ be the number of reported vertices. The total reporting time is bounded by $\sum_{Q \in S(b)}k_Q = O(k \log n)$.

Using the results of Section \ref{sec:ds:path-tree} we can get join-reachability structures when $G_2$ is a rooted or an unoriented tree. Let $I_2(a)$ be the depth-first search interval assigned to each vertex $a$ in $T_2$, where $T_2$ is the undirected version of $G_2$. If $G_2$ is an out-tree then we report the vertices $a$ that satisfy  $\mathrm{to}_a[Q] \le \mathrm{from}_b[Q]$ and $I_2(b) \subseteq I_2(a)$, which by Section \ref{sec:ds:path-tree} can be done in $O(k_Q)$ time. So, the total reporting time is $O(k \log n)$. Similarly, if $G_2$ is an in-tree then we report the vertices $a$ that satisfy  $\mathrm{to}_a[Q] \le \mathrm{from}_b[Q]$ and $I_2(a) \subseteq I_2(b)$, which again takes $O(k_Q)$ time with the structure of Section \ref{sec:ds:path-tree}. So, the total reporting time in both cases is bounded by $O(k \log{n})$. Theorem \ref{thm:ds}(d) follows.
With similar ideas we can obtain an $\langle n \log^2{n},  k \log^2{n} \rangle$ structure when $G_2$ is also a planar digraph, as stated by Theorem \ref{thm:ds}(e).

\subsection{General Digraphs}
\label{sec:general}

Here we examine how to obtain join-reachability strucures for general digraphs with the use of dipath covers. We begin with the case where $G_2$ is a dipath.

Let $P^{1}_1, P^{2}_{1}, \ldots P^{\kappa_1}_{1}$ be a dipath cover of $G_1$, and let
$P^{i}_{2}$ be the dipath that consists of the vertices in $P^{i}_{1}$ ordered by increasing rank in $G_2$.
Also, let $V_{P^{i}_{1}}$ be set of vertices that have a predecessor in $P^{i}_{1}$. We build a join-reachability structure for each pair $\{P^{i}_{1},P^{i}_{2}\}$ which we use in order to report the vertices in $P^{i}_{1}$ that reach a query vertex in both $G_1$ and $G_2$. To that end, each vertex $a$ in $P^{i}_1$ is assigned coordinates $x_1(a) = r_{P_{1}^{i}}(a)$ and $x_2(a) = r_{P_{2}^{i}}(a)$, and we build a join-reachability structure for these vertices as in Section \ref{sec:ds:paths}. With this structure we can answer a reporting query for vertex $b$ by finding the vertices $a$ that satisfy $(x_1(a),x_2(a)) \le (\mathrm{from}_b[P_{1}^{i}],\mathrm{from}_b[P_{2}^{i}])$ for each $i \in \{1,\ldots,\kappa_1\}$. The reporting time is $O(k+\kappa_1)$ using $O(\kappa_1 n)$ space. The reporting time can be reduced to $O(k)$ if we store for each vertex $v$ a list $I(v)$ of the indices $i \in \{1,\ldots,\kappa_1\}$ such that the reporting query for $v$ in the join-reachability structure for the pair $\{ P_{1}^{i}, P_{2}^{i} \}$ is non-empty. Then we only need to query the structures for $i \in I(v)$. The asymptotic space bound remains $O(\kappa_1 n)$.

We can extend the above method in order to handle two general graphs. The resulting bounds, however, are interesting only when the product $\kappa_1 \kappa_2$ is small compared to $n$, where $\kappa_2$ is the number of disjoint dipaths in a dipath cover of $G_2$. Specifically, we can get either $O((\kappa_1+\kappa_2)n)$ space and $O(\kappa_1 \kappa_2 + k)$ reporting time, or $O((\kappa_1\kappa_2)n)$ space and $O(k)$ reporting time. (In the latter structure we improve the reporting time by storing for each vertex $v$ the pairs of dipaths in the cover of $G_1$ and $G_2$ that contain a common predecessor of $v$.) This implies Theorem \ref{thm:ds}(h).
By combining the dipath cover method with the techniques of Section \ref{sec:ds:path-tree} we obtain the bound of Theorem \ref{thm:ds}(f).
Similarly, the techniques of Section \ref{sec:ds:planar} imply Theorem \ref{thm:ds}(g).

\section{Conclusions and Open Problems}
\label{sec:conclude}

We explored the computational and combinatorial complexity of the join-reachability graph, and the design of efficient join-reachability data structures for a variety of graph classes. We believe that several open problems deserve further investigation. For instance, from the aspect of combinatorial complexity, it would be interesting to prove or disprove that an $O(m \cdot \mathrm{polylog}(n))$ bound on the size of the join-reachability graph $\mathcal{J}(\{G_1,G_2\})$ is attainable when $G_1$ is a general digraph with $n$ vertices and $m$ arcs and $G_2$ is a dipath. Another direction is to consider the problem of approximating the smallest join-reachability graph for specific graph classes. From the aspect of data structures, we can consider the following type of join-reachability query: Given vertices $b$ and $c$, report (or count) all vertices $a$ such that $a \reach_{G_1} b$ and $a \reach_{G_2} c$.

\paragraph{Acknowledgement.} We would like to thank Li Zhang for several useful discussions.

\bibliographystyle{plain}
\bibliography{join-reachability}

\begin{thebibliography}{10}

\bibitem{transitive-db:abj}
R.~Agrawal, A.~Borgida, and H.~V. Jagadish.
\newblock Efficient management of transitive relationships in large data and
  knowledge bases.
\newblock In {\em SIGMOD '89: Proceedings of the 1989 ACM SIGMOD international
  conference on Management of data}, pages 253--262, 1989.

\bibitem{transitive-reduction:agu}
A.~V. Aho, M.~R. Garey, and J.~D. Ullman.
\newblock The transitive reduction of a directed graph.
\newblock {\em SIAM J. Comput.}, 1(2):131--137, 1972.

\bibitem{range-searching:abr}
S.~Alstrup, G.~S. Brodal, and T.~Rauhe.
\newblock New data structures for orthogonal range searching.
\newblock In {\em FOCS '00: Proceedings of the 41st Annual Symposium on
  Foundations of Computer Science}, page 198, 2000.

\bibitem{filt:c86}
B.~Chazelle.
\newblock Filtering search: {A} new approach to query-answering.
\newblock {\em SIAM Journal on Computing}, 15(3):703--24, 1986.

\bibitem{rank-aggregation:ckns01}
C.~Dwork, R.~Kumar, M.~Naor, and D.~Sivakumar.
\newblock Rank aggregation methods for the web.
\newblock In {\em WWW '01: Proceedings of the 10th international conference on
  World Wide Web}, pages 613--622, 2001.

\bibitem{scaling:gbt84}
H.~N. Gabow, J.~L. Bentley, and R.~E. Tarjan.
\newblock Scaling and related techniques for geometry problems.
\newblock In {\em Proc. 16th ACM Symp. on Theory of Computing}, pages 135--143,
  1984.

\bibitem{fdom:G}
L.~Georgiadis.
\newblock Computing frequency dominators and related problems.
\newblock In {\em ISAAC '08: Proceedings of the 19th International Symposium on
  Algorithms and Computation}, pages 704--715, 2008.

\bibitem{2-vc:g}
L.~Georgiadis.
\newblock Testing $2$-vertex connectivity and computing pairs of
  vertex-disjoint $s$-$t$ paths in digraphs.
\newblock In {\em Proc. 37th Int'l. Coll. on Automata, Languages, and
  Programming}, pages 738--749, 2010.

\bibitem{domv:gt05}
L.~Georgiadis and R.~E. Tarjan.
\newblock Dominator tree verification and vertex-disjoint paths.
\newblock In {\em Proc. 16th ACM-SIAM Symp. on Discrete Algorithms}, pages
  433--442, 2005.

\bibitem{nca:ht}
D.~Harel and R.~E. Tarjan.
\newblock Fast algorithms for finding nearest common ancestors.
\newblock {\em SIAM Journal on Computing}, 13(2):338--55, 1984.

\bibitem{multidim-dom:jms04}
J.~Ja{J}a, C.~W. Mortensen, and Q.~Shi.
\newblock Space-efficient and fast algorithms for multidimensional dominance
  reporting and counting.
\newblock In {\em ISAAC '04: Proceedings of the 15th International Symposium on
  Algorithms and Computation}, pages 558--568, 2004.

\bibitem{planar-reach:kameda}
T.~Kameda.
\newblock On the vector representation of the reachability in planar directed
  graphs.
\newblock {\em Information Processing Letters}, 3(3):75--77, 1975.

\bibitem{reach-sub:KKS}
I.~Katriel, M.~Kutz, and M.~Skutella.
\newblock Reachability substitutes for planar digraphs.
\newblock Technical Report MPI-I-2005-1-002, Max-Planck-Institut F\"{u}r
  Informatik, 2005.

\bibitem{persistent:st}
N.~Sarnak and R.~E. Tarjan.
\newblock Planar point location using persistent search trees.
\newblock {\em Communications of the ACM}, 29(7):669--679, 1986.

\bibitem{dyntrees:st83}
D.~D. Sleator and R.~E. Tarjan.
\newblock A data structure for dynamic trees.
\newblock {\em Journal of Computer and System Sciences}, 26:362--391, 1983.

\bibitem{lattice-reach:tv99}
M.~Talamo and P.~Vocca.
\newblock An efficient data structure for lattice operations.
\newblock {\em SIAM J. Comput.}, 28(5):1783--1805, 1999.

\bibitem{planar-reach:tt}
R.~Tamassia and I.~G. Tollis.
\newblock Dynamic reachability in planar digraphs with one source and one sink.
\newblock {\em Theoretical Computer Science}, 119(2):331--343, 1993.

\bibitem{planar-reach:thorup}
M.~Thorup.
\newblock Compact oracles for reachability and approximate distances in planar
  digraphs.
\newblock {\em Journal of the ACM}, 51(6):993--1024, 2004.

\bibitem{rout:tz}
M.~Thorup and U.~Zwick.
\newblock Compact routing schemes.
\newblock In {\em Proc. 13th ACM Symp. on Parallel Algorithms and
  Architecture}, pages 1--10, 2001.

\bibitem{reachability:WHYYY06}
H.~Wang, H.~He, J.~Yang, P.~S. Yu, and J.~X. Yu.
\newblock Dual labeling: Answering graph reachability queries in constant time.
\newblock In {\em ICDE '06: Proceedings of the 22nd International Conference on
  Data Engineering}, page~75, 2006.

\end{thebibliography}

\clearpage

\begin{appendices}

In the Appendices we provide additional join-reachability data structures. In Appendix \ref{appendix:path-decomposition} we apply the \emph{heavy-path decomposition} of trees~\cite{dyntrees:st83} in order to get alternative join-reachability data structures for trees and paths. In Appendix \ref{appendix:planar-st} we consider the case of planar $st$-graphs~\cite{planar-reach:tt}, and in Appendix \ref{appendix:lattice} we consider lattices.

\section{Join-Reachability for Trees based on Heavy-Path Decomposition}
\label{appendix:path-decomposition}

Let $T$ be the rooted tree that results from $G_1$ after removing arc directions.
We develop a method based on partitioning $T$ into \emph{heavy paths}~\cite{dyntrees:st83}. This is done as follows. A child $a'$ of $a$ is \emph{heavy} if $|T(a')| \ge |T(a)|/2$, and \emph{light} otherwise. The \emph{light level} of a vertex $a$ is the number of light vertices on the path from $a$ to the root of $T$. Each vertex has at most one heavy child and its light level is $O(\log n)$. The heavy paths are formed by the edges connecting a heavy child to its parent and the topmost vertex of a heavy path is light.

First we consider the case where $G_2$ is a dipath, and then the case where $G_2$ is a (in- or out-)tree.

\subsection{Tree and Path}
\label{appendix:path-decomposition:tree-path}

Based on the heavy-path decomposition of $T$, we describe a structure with $O(k \log n)$ reporting time for an in-tree and $O(\log{n}+k)$ reporting time for an out-tree. These bounds are inferior to the ones given in Section \ref{sec:ds:path-tree}, but are achieved with simpler structures.

Consider the in-tree query first. Here our method is inspired by a routing scheme for trees by Thorup and Zwick~\cite{rout:tz}.
Let $h(T(a))$ be the maximum label in $T(a)$. Obviously, we need to search $T(a)$ only if $h(T(a)) > j$.
Let $h'(T(a))$ be the maximum label in $T(a) \setminus T(a')$, where $a'$ is the heavy child of $a$ (if it exists). The search proceeds top-down starting from $b$. Let $a$ be the current vertex such that $h(T(a)) > j$. If $h(a) > j$, we report $a$. Then we identify the light children $c$ of $a$ such that $h(T(c)) > j$. Moreover, if $a$ is the topmost vertex of its heavy path $P$, then we identify the vertices $d \in P$ such that $h'(T(d)) > j$. Then, we repeat this process at each vertex that we have identified. In order to locate these vertices quickly, for each vertex $a$ we order its light children $c$ by $h(T(c))$, and for each heavy path $P$ we order the vertices $d \in P$ by $h'(T(d))$. Note that when we visit a light child $c$, the light level increases and there is at least one $x \in T(c)$ with $h(x)>j$. The $O(k \log n)$ bound follows.

For the out-tree query we use the same heavy-path decomposition and construct a Cartesian tree for each heavy path $P$. (See Section ~\ref{sec:ds:paths}). The Cartesian tree for $P$ stores the vertices in $a \in P$ according to coordinates $(x_1(a),x_2(a))=(h_{P}(a),h_{G_2}(a))$. Furthermore, each vertex has a pointer to the topmost vertex of its heavy path, and each topmost vertex of a heavy path has a pointer to its parent in $T$. Let $b$ be the query vertex and let $Q$ be the tree path from the root of $T$ to $b$. The goal is to identify the vertices $a \in Q$ with $h(a)>j$. We locate the heavy paths that intersect $Q$ and query them individually. For each such heavy path $P$ we identify the bottommost vertex $p \in P \cap Q$.
The query for $P$ has to report the vertices $a \in P$ such that $(x_1(a),x_2(a)) \ge (x_1(p),j)$. As mentioned in Section~\ref{sec:ds:paths}, Cartesian trees can report these vertices in constant time per vertex. Since $Q$ intersects $O(\log n)$ heavy paths the total query time is $O(\log n + k)$.

\subsection{Two Trees}
\label{appendix:path-decomposition:two-trees}

With the heavy-path decomposition method we can get an efficient join-reachability structure when one of the two trees is an out-tree. Without loss of generality we assume that $G_1$ is an out-tree. We perform the heavy-path decomposition of $T$ as earlier and associate with each heavy path $P$ a secondary data structure $D_P$; the choice of the secondary structure depends on the type of $G_2$. Also for each vertex $a \in P$ we store $h_P(a)$, the height of $a$ in $P$.
Given a query vertex $b$ we want to report the ancestors $a$ of $b$ in $T$ that reach $b$ in $G_2$. Let $Q$ be the path in $T$ from the root to $b$. Our algorithm queries the structure $D_P$ for each heavy path $P$ that intersects $Q$. For each such heavy path $P$ we identify the bottommost vertex $p \in P \cap Q$.
If $G_2$ is an out-tree then we need to report the vertices $a \in P$ that satisfy $I_2(b) \subseteq I_2(a)$ and $h_P(a) \ge h_P(p)$. In this case, a suitable choice for $D_P$ is a join-reachability structure for an out-tree and a path. Either of the two solutions we developed earlier (Sections \ref{sec:ds:path-tree} and \ref{appendix:path-decomposition:tree-path}) achieves $O(\log{|P|}+k_P)$ reporting time (because here we need to locate $b$ in $D_P$), where $k_P$ is the number of reported vertices on $P$. This results to an overall $\langle n,  \log^2{n} + k \rangle$ structure.
For the case where $G_2$ is an in-tree we need to report the vertices $a \in P$ that satisfy $I_2(a) \subseteq I_2(b)$ and $h_P(a) \ge h_P(p)$.
Here we choose $D_P$ to be a join-reachability structure for an in-tree and a path. Using the geometry-based structure of Section \ref{sec:ds:path-tree} results to an overall $\langle n,  \log^2{n} + k \rangle$ structure.

\section{Planar $st$-Graphs}
\label{appendix:planar-st}

Here we consider the case where $G_1$ is a \emph{planar $st$-graph}~\cite{planar-reach:tt} and $G_2$ is a dipath. A planar $st$-graph is planar acyclic digraph with a single source $s$ and a single sink $t$, such that $s$ and $t$ are on the boundary of the same face. For these graphs Kameda~\cite{planar-reach:kameda} gave an $O(n)$-space structure that answers reachability queries in constant time. His algorithm performs two modified depth-first searches and assigns to each vertex $a$ two integer labels $\ell_1(a)$ and $\ell_2(a)$ both in the range $[1,n]$. Kameda then shows that these labels satisfy the property that
 $a \rightsquigarrow_{G_1} b$ if and only if $\ell_1(a) \le \ell_1(b)$ and $\ell_2(a) \le \ell_2(b)$.
Our data structure for the join-reachability problem also assigns each vertex $a$ a third label $\ell_3(a)$ equal to the rank of $a$ in $G_3$. Now each vertex corresponds to a point in a three-dimensional rank space and  $a \rightsquigarrow_{\mathcal{J}} b$ if and only if $(\ell_1(a),\ell_2(a),\ell_3(a)) \le (\ell_1(b),\ell_2(b),\ell_3(b))$. Using a three-dimensional dominance structure we can get an $\langle n , \log{n}+k \rangle$ join-reachability structure~\cite{multidim-dom:jms04}.
With minor adjustments we can get an efficient data structure for the more general class of \emph{spherical $st$-graphs}~\cite{planar-reach:tt}, which are planar $st$-graphs without the requirement that $s$ and $t$ appear on the boundary of the same face. Tamassia and Tollis~\cite{planar-reach:tt} showed how to reduce the reachability problem on these graphs to a reachability problem on planar $st$-graphs.

\section{Lattices}
\label{appendix:lattice}

Let $(\le,V)$ be a partial order. An element $z \in V$ is an \emph{upper bound} of $x ,y \in V$ if $x \le z$ and $y \le z$. If $z$ is an upper bound of $x,y$ and moreover $z \le w$ for all upper bounds $w$ of $x,y$ then $z$ is a \emph{least upper bound} of $x,y$. Similarly, if $z \le x$ and $z \le y$ then $z$ is a \emph{lower bound} of $x,y$, and if $w \le z$ for all lower bounds $w$ of $x,y$ then $z$ is a \emph{greatest lower bound} of $x,y$. A partial order $(\le,V)$ is a \emph{lattice} if any two $x,y \in V$ have both a least upper bound and a greatest lower bound. A \emph{partial lattice} $(\le,V)$ is a partial order that can be extended to lattice by adding elements $s$ and $t$ such that $s \le x$ and $x \le t$ for any $x \in V$. Any acyclic digraph $G=(V,A)$ has an associated partial order $P_G=(\le,V)$ such that for $u, v \in V$, $u \le v$ if and only if $u \reach_G v$. We say that $G$ \emph{satisfies the lattice property} if and only if its associated partial order is a lattice. For this class of digraphs, Talamo and Vocca presented an $O(n \sqrt{n})$-space structure that answers reachability queries in constant time~\cite{lattice-reach:tv99}. Their structure is also capable of reporting the predecessors of a query vertex in $O(k)$ time. In this section we show how their structure can be extended in order to  support efficient join-reachability.
Roughly speaking, the Talamo-Vocca structure represents $G$ as a collection of disjoint clusters with $O(\sqrt{n})$ vertices each. Moreover, we can assume that there are $\Theta(\sqrt{n})$ clusters; refer to \cite{lattice-reach:tv99} for details. Each cluster $C$ has a root vertex $c$ and consists of either a subset of the predecessors of $c$, in which case it is an \emph{in-cluster}, or of a subset of the successors of $c$, in which case it is an \emph{out-cluster}. A vertex $x \in C$ is an \emph{internal vertex of} $C$; a vertex $x \not\in C$ that either reaches or is reachable from a vertex in $C$ is an \emph{external vertex of} $C$. External vertices have the following key property: If $x$ is an external vertex that reaches (resp. is reachable from) a subset $S \subseteq C$ then $S$ contains the greatest lower bound (resp. least upper bound) of $S$, which is the \emph{representative of }$x$\emph{ in }$C$. Now each vertex $x$ is associated with a subgraph $G(x)$ consisting of two trees rooted at $x$; an internal spanning tree $I(x)$ and an external spanning tree $E(x)$. If the cluster $C$ containing $x$ is an in-cluster then the internal tree is an in-tree that contains the predecessors of $x$ in $C$ and the external tree is an out-tree that contains the external vertices of $C$ with $x$ as their representative. Similarly, if the cluster $C$ containing $x$ is an out-cluster then the internal tree is an in-tree that contains the successors of $x$ in $C$ and the external tree is an in-tree that contains the external vertices of $C$ with $x$ as their representative. In order to be able to report all the predecessors of a query vertex $b$ this data structure can explicitly store the predecessors of each vertex $x$ that are located in the same cluster with $x$. Since each cluster has $O(\sqrt{n})$ vertices the data structure still occupies $O(n\sqrt{n})$ space. The predecessors of $b$ outside its cluster are the predecessors of the vertices that are representatives of $b$ in other clusters for which $b$ is an external vertex.

We can easily enhance the above structure so that it supports efficient join-reachability. We demonstrate this first for the case where $G_2$ is a dipath. For each vertex $x$ we construct a list $L_1(x)$ of the internal predecessors of $x$ sorted in increasing rank in $G_2$. Also we keep track of the minimum rank in $G_2$ of the vertices in $L_1(x)$. Then we construct another list $L_2(x)$ which contains the representatives of $x$ in the clusters where $x$ is an external vertex. Furthermore, $y \in L_2(x)$ only if the minimum rank in $L_1(y)$ is less than the rank of $x$. Now in order to report the vertices reaching $b$ in the join-reachability graph, we report the vertices in $L_1(a)$ with rank less than $b$, for all $a \in L_2(b) \cup \{ b \}$. Notice that we only visit clusters that contain a least one predecessor of $b$. Therefore the reporting time is $O(k)$.

Now we show how the same bounds are achieved when $G_2$ is a rooted tree. Let $I_2(a)=[s_2(a),t_2(a)]$ be the depth-first search interval assigned to each vertex $a$ in $T_2$, where $T_2$ is the undirected version of $G_2$. For each vertex $x$ we construct a structure $D(x)$ that contains the vertices in $L_1(x)$. In order to report the vertices that reach $b$ in the join-reachability graph, we query the structure $D(a)$ for all $a \in L_2(b) \cup \{ b \}$. This structure reports the vertices $\gamma \in L_1(a)$ that satisfy $I_2(b) \subseteq I_2(\gamma)$ if $G_2$ is an out-tree, or $I_2(\gamma) \subseteq I_2(b)$ if $G_2$ is an in-tree. Note that during the construction of the join-reachability data structure we can ensure that $a \in L_2(b) \cup \{ b \}$ only if the answer of $D(a)$ to query $b$ is nonempty. Finally, we need to specify how $D(a)$ operates. If $G_2$ is an out-tree then $D(a)$ stores the intervals $I_2(\gamma)$ for all $\gamma \in L_1(a)$;
a query asks for those $\gamma \in L_1(a)$ such that $I_2(\gamma)$ contains the point $s_2(b)$. Otherwise, when $G_2$ is an in-tree, $D(a)$ stores the points $s_2(\gamma)$ for all $\gamma \in L_1(a)$; now a query asks for those $\gamma \in L_1(a)$ such that $s_2(\gamma)$ is contained in $I_2(b)$. Such queries can be answered optimally by Chazelle's \emph{interval overlap} structure~\cite{filt:c86}, which gives us the desired result.

\begin{theorem}
\label{thm:lattice}
Given a lattice $G_1$ and an unoriented tree $G_2$ with $n$ vertices we can construct an $\langle n \sqrt{n}, k \rangle$ join-reachability data structure.
\ignore{Also the size of the join-reachability graph in this case is $O(n \sqrt{n})$.}
\end{theorem}

\end{appendices}

\end{document}